\documentclass[
  journal=pasa,
  manuscript=research-paper, 
  year=2025,
  volume=42,
]{cup-journal}
\usepackage{graphicx}
\usepackage{dcolumn}
\usepackage{bm}
\usepackage{multirow}
\usepackage{threeparttable}
\usepackage{adjustbox}
\usepackage{rotating}
\usepackage{caption}
\usepackage{subfigure}
\usepackage{subcaption}
\usepackage{hyperref}
\usepackage{float}
\usepackage{footnote}
\usepackage{microtype,siunitx,booktabs}
\usepackage{amsmath,amssymb}
\usepackage{xcolor}
\hypersetup{colorlinks=true,citecolor=blue,linkcolor=blue,urlcolor=blue}
\usepackage{orcidlink}
\DeclareUnicodeCharacter{2212}{-}

\title{Post-glitch Recovery and the Neutron Star Structure: The Vela Pulsar}

\author{Himanshu Grover\,\orcidlink{0009-0004-1150-6151}}
\affiliation{Department of Physics, Indian Institute of Technology Roorkee, Roorkee 247667, India}
\email[H. Grover]{himanshu\_g@ph.iitr.ac.in}

\author{Erbil Gügercinoğlu}
\affiliation{School of Arts and Sciences, Qingdao Binhai University, Huangdao District, Qingdao, 266555, China} 
\alsoaffiliation{National Astronomical Observatories, Chinese Academy of Sciences, 20A Datun Road, Chaoyang District, Beijing, 100101, China}

\author{Bhal Chandra Joshi\,\orcidlink{0000-0002-0863-7781}}
\affiliation{National Centre for Radio Astrophysics, TIFR, P Bag 3, Ganeshkhind, Pune - 411007 India}
\alsoaffiliation{Department of Physics, Indian Institute of Technology Roorkee, Roorkee 247667, India}

\author{M.A. Krishnakumar\,\orcidlink{0000-0003-4528-2745}}
\affiliation{National Centre for Radio Astrophysics, TIFR, P Bag 3, Ganeshkhind, Pune - 411007 India}

\author{Shantanu Desai\,\orcidlink{0000-0002-0466-3288}}
\affiliation{Department of Physics, Indian Institute of Technology Hyderabad, Kandi, Telangana 502284, India}   

\author{P. Arumugam\,\orcidlink{0000-0001-9624-8024}}
\affiliation{Department of Physics, Indian Institute of Technology Roorkee, Roorkee 247667, India}	
        
\author{Debades  Bandyopadhyay\,\orcidlink{0000-0003-0616-4367}}
\affiliation{Saha Institute of Nuclear Physics, 1/AF Bidhannagar, Kolkata - 700064, India}
        
\doi{xx.yyyy/pasa.zzzz.tt}

\received {dd Mmm YYYY}
\revised  {dd Mmm YYYY}
\accepted {dd Mmm YYYY}
\published{dd Mmm YYYY}


\begin{document}

\begin{abstract}
We present a detailed analysis of the Vela pulsar's rotational behaviour using approximately 100 months of observational data spanning from September 2016 to January 2025, during which four glitches were identified. Here, we demonstrate the post-glitch recovery of these glitches within the framework of the vortex creep model. We further present the investigation of vortex residuals (the discrepancy between observed values and those predicted by the vortex creep model) by interpreting them in the context of the vortex bending model. In addition, we report a positive correlation between the glitch magnitude and the time to the next glitch, applicable only for the large glitch events observed in the Vela pulsar. Furthermore, we estimate the braking index of the Vela pulsar to be 2.94 $\pm$ 0.55.
\end{abstract}

\maketitle

\section{Introduction}
\label{introduction}
Pulsars are highly magnetised, rotating, dense, and compact stars observed to have periodic radiation of pulses with the periodicity tied to their remarkably stable rotation. The slowdown in the rotation of a pulsar is influenced by two types of instabilities: glitches and timing noise \citep{D'alessandro96, Haskell_2015, zhou_erbil_glitch_review}. Glitches are abrupt changes in the rotational frequency and spin-down rate of a pulsar, while timing noise is a stochastic wander in the spin parameters of pulsars. The occurrence of glitches, as well as their slow relaxations, which range from a few days to several years, reveals the presence of neutron superfluidity within the inner crust of the neutron star. One of the most widely accepted mechanisms for explaining glitches and their recoveries is the superfluid vortex pinning and creep model \citep{AndersonItoh1975, Alpar1984I, Alpar1984II}. This model suggests that neutron superfluidity in the interior of the star leads to a differential rotation, storing a significant amount of angular momentum. When this angular momentum is released, the crust spins up, resulting in a glitch. Pulsar glitches thus provide valuable insights into the superfluid dynamics within the star \citep{zhou_erbil_glitch_review, Liu2024}. 

The post-glitch recovery is the process in which the pulsar gradually relaxes back towards its original pre-glitch rotational state. According to the vortex creep model, this process is driven by the repinning of superfluid vortices to the crustal nuclei and the re-establishment of the equilibrium lag between the rotational rates of the superfluid interior and the crust. The specific timescale and form of the relaxation, whether exponential or linear, are determined by the distinct vortex dynamics at play. While the standard recovery reflects the transfer of angular momentum from the interior to the surface, deviations from this norm can indicate complex interactions, such as crustal events or non-linear creep regimes \citep{Akbal_2015, Erbil_2023}. Ultimately, the post-glitch recovery serves as a probe for the internal dynamics and structure of neutron stars \citep{Erbil_2022}.

PSR J0835--4510, B0833--45, the Vela pulsar, has 26 reported glitch events, most of which are large glitches and detected after a regular, quasi-periodic inter-glitch time\footnote{\url{http://www.jb.man.ac.uk/pulsar/glitches.html}} \citep{Espinoza2011, Basu2022}. The Vela pulsar is one of the most studied glitching pulsars. The first-ever glitch was recorded in the Vela pulsar \citep{Radhakrishnan_Manchester_1969, Reichley_Downs_1969}. Moreover, the first-ever glitch detection caught in the act was also realised for the Vela pulsar \citep{Palfreyman2018_Vela}.
The Vela pulsar is considered one of the most popular glitching pulsars for several reasons: it is a bright, young source that displays glitches at regular intervals. The recoveries of glitches also demonstrate strikingly similar behaviour. A measure of a pulsar's slowdown over time is described by the braking index ($n$). It is defined by a power-law relation between the pulsar's spin-down rate $\dot{\Omega}$ and its rotational angular frequency ($\Omega$), given by $\dot{\Omega} = k \Omega^n$. The braking index provides insights into the pulsar's rotational evolution, magnetic fields, and emission mechanisms. 

Here, we present the observational behaviour of the Vela pulsar from September 2016 to January 2025; during this period, four large glitches were observed in our monitoring campaign \citep{Grover2024, Basu2020}. Theoretical inferences from the post-glitch recovery phase are also presented. We have used the recovery behaviour of four glitches in the Vela pulsar to estimate the following parameters: (a) the fractional moments of inertia (FMI) of various layers participating in a glitch, (b) the re-coupling timescale of the crustal superfluid, (c) theoretical prediction for the time to the next glitch, (d) braking index of the Vela pulsar. The paper's outline is as follows: a brief description of the vortex creep and bending models is given in Section \ref{vortexCreepModel}. We describe the pulsar observations, the data reduction, and the analysis processes in section \ref{PulsarObservations}. In Section \ref{ResultsandDiscussions}, we present the results. The conclusions and future scope are given in Section \ref{conclusion}.

\section{The Vortex Creep and Bending Models}\label{vortexCreepModel}
The vortex creep model provides a microscopic framework for understanding the macroscopic behaviour of glitches and their subsequent recovery. In the neutron star crust, vortex lines coexist with and energetically pin to the crystal lattice. Crustal inhomogeneities such as dislocations, cracks, and impurities produced by quakes create a non-uniform vortex distribution known as vortex traps \citep{Cheng1988}. Glitches may arise due to collective vortex unpinning events induced in these vortex traps when the angular velocity lag between the superfluid and the crustal normal matter exceeds a threshold value \citep{AndersonItoh1975, Alpar1984I}. During the unpinning events, angular momentum is rapidly transferred from the superfluid to the crust. The superfluid regions involved in the unpinning decouple from the external braking torque, resulting in a step increase in the spin-down rate. 

After a glitch, the steady-state lag decreases, temporarily suppressing creep in some superfluid regions. The recovery process involves restoring the creep until steady-state conditions are re-established by the external braking torque. Shortly after the glitch, excess angular momentum is transferred from the crustal superfluid to the normal matter in the crust, producing a transient peak in the crustal rotation rate. This perturbation is subsequently communicated to the core superfluid through electron scattering off magnetised vortex lines \citep{Alpar1984II}, causing the initial peak to relax on the crust–core coupling timescale toward a new equilibrium state. A pronounced peak in the rotation rate of the normal component is expected when the coupling between the crustal superfluid and the normal matter is stronger than that with the core superfluid \citep{Pizzochero2020}. On longer timescales, the core superfluid recouples, leaving only the crustal superfluid effectively decoupled from the external braking torque.

In some glitches of the Crab pulsar, delayed spin-ups are observed \citep{Wong2001, Shaw2018, Basu2020, Ge_2020}. Within the vortex creep framework, these features are attributed to vortex unpinning accompanied by inward vortex motion driven by fractured crustal platelets \citep{Akbal_2015, Erbil_2019}. Additionally, damped, sinusoidal-like oscillations in the spin-down rate can arise from vortex-line bending in the non-linear creep relaxation regime, excited by the glitch and potentially influenced by preceding events \citep{Erbil_2023}. It is important to note that the vortex creep model is not the unique explanation; several equally viable models successfully explain glitch-triggering mechanisms, delayed spin-ups, and post-glitch relaxation. For a comprehensive discussion, see \cite{Haskell_2015, Antonopoulou2022, Antonelli2022, zhou_erbil_glitch_review}.

The post-glitch spin-down rate behaviour mirrors the responses of the vortex creep regions in both linear and non-linear regimes to a glitch, with additional involvement of induced inward vortex motion in delayed spin-up events. In certain regions of the superfluid in the inner crust and the outer core, vortex creep exhibits a linear response to glitch-induced changes, leading to exponential relaxation \citep{Alpar1989, Flanagan1990, Erbil_2017aug} as:
\begin{equation}\label{eqn_exp_recovery}
    \Delta \Dot{\nu}(t) = \sum_{i=1}^{3} - \frac{I_{\text{ei}}}{I_{\text{c}}} \frac{\Delta \nu}{\tau_{\text{ei}}} e^{-t \slash  \tau_{\text{ei}}},
\end{equation}
where $\Delta \nu = \nu - \nu_0$, $\nu$ is the spin frequency, $\nu_0$ is the spin frequency just before the glitch, $I_\text{ei}\slash I_\text{c}$ is the FMI of the of i$^{th}$ layer that participated in a glitch, and $\tau_\text{ei}$ is the corresponding exponential recovery timescale.

The overall linear relaxation, obtained by integrating the contributions of several neighbouring single non-linear regime regions \citep{Alpar1984I, Erbil_2022}, with an assumption of linearly decreasing superfluid angular velocity during a glitch \citep{Alpar1996} and is given as:
\begin{equation}\label{eqn_linear_recovery}
    \Delta \Dot{\nu}(t) = \frac{I_{\text{a}}}{I_{\text{c}}} \Dot{\nu}_0 \left[1 - \frac{1- (\tau_\text{nl} \slash t_0) \text{ln}[1+(\text{e}^{t_0\slash \tau_\text{nl}} - 1)\text{e}^{-t\slash \tau_\text{nl}}]}{1-\text{e}^{-t\slash \tau_\text{nl}}} \right],
\end{equation}
where $I_\text{a}\slash I_\text{c}$ is the FMI of the non-linear crustal superfluid region, $t_0$ is the offset time, and $\tau_\text{nl}$ is the non-linear creep relaxation time. The non-linear creep response gives an estimate for the time to the next glitch. The glitch-induced changes in the spin-down rate fully relax after a waiting time \citep{Erbil_2022},
\begin{equation}\label{eqn_waiting_time}
    t_0 = \frac{\delta\nu_\text{s} + \Delta\nu_\text{c}}{|\Dot{\nu}|} \cong \frac{\delta\nu_\text{s}}{|\Dot{\nu}|}.
\end{equation}
where $\Delta\nu_\text{c}$ is the observed glitch magnitude and $\delta\nu_\text{s}$ represents the decrement in the superfluid rotational frequency. The post-glitch recovery completes once the relaxation of the non-linear creep regions is finished. This gives us a theoretical prediction for the time to the next glitch in terms of the vortex creep model, which lies between the range $t_0-3\tau_{nl}$ to $t_0+3\tau_{nl}$.

A typical post-glitch recovery involves an exponential and a linear relaxation described by equations~\ref{eqn_exp_recovery} and \ref{eqn_linear_recovery}, respectively. This is tested by using different alternative models as described below:

\begin{itemize}
    \item Model 1: Only exponential recovery given by equation~\ref{eqn_exp_recovery}.
    \begin{itemize}
        \item Model 1a: Exponential recovery with only 1 exponential component.
        \item Model 1b: Exponential recovery with 2 exponential components.
        \item Model 1c: Exponential recovery with 3 exponential components.
    \end{itemize}
    \item Model 2: Only linear relaxation described by equation~\ref{eqn_linear_recovery}.
    \item Model 3: Exponential and linear recoveries, the sum of equations~\ref{eqn_exp_recovery} and~\ref{eqn_linear_recovery}.
    \begin{itemize}
        \item Model 3a: The sum of linear recovery and exponential recovery with only 1 exponential component.
        \item Model 3b: The sum of linear recovery and exponential recovery with 2 exponential components.
        \item Model 3c: The sum of linear recovery and exponential recovery with 3 exponential components.
    \end{itemize}
\end{itemize}    

The most preferred model and corresponding parameters among the various alternatives were evaluated using Bayesian analysis of our post-glitch recovery measurements of the Vela pulsar. Further details of this analysis are presented in the next section.

Apart from long-term post-glitch recovery, we see evidence for sinusoidal-like spin-down rate oscillations over and above the recovery predicted by the vortex creep model as described in Section~\ref{ResultsandDiscussions}. Such oscillations can be explained by vortex bending \citep{Erbil_2023} and appear to be a common feature following glitches, as also observed from PSRs J1522--5735 \citep{zhou24}, J0742--2822 \citep{zubieta25}, J1509--5850, and J1718--3825 \citep{liu25}. To understand this better, we used three alternative hypotheses as described below

\begin{itemize}
    \item Hypothesis I -- An undamped model consisting of only sinusoidal term, $A\sin (\omega t+\phi)$.
    \item Hypothesis II -- Damped model consisting of a sinusoidal term and a damping factor, $A\sin (\omega t+\phi) \exp(-t/\tau)$.
    \item Hypothesis III -- An oscillatory model as described in \cite{Erbil_2023} consisting of $A\omega\cos(\omega t+\phi)\exp(-t/2\tau)-(A/2\tau) \sin(\omega t+\phi) \exp(-t/2\tau)$.
\end{itemize}

where $A$ is the amplitude, $\omega$ is the angular frequency, $\phi$ is the phase, and $\tau$ is the decay time for the oscillation. The next section outlines the observations, data reduction, and timing used to derive the spin frequency and spin-down rate. It also describes the Bayesian analysis used to calculate the evidence for the competing hypotheses, thereby identifying the most plausible model and estimating its corresponding parameters. 

\section{Observations and Analysis}\label{PulsarObservations}

\subsection{Sample and Observational Facilities}
We use two large radio telescopes to monitor pulsars: the upgraded Giant Metrewave Radio Telescope (uGMRT)\citep{ugmrt2017} and the Ooty Radio Telescope (ORT) \citep{swarup1971ort}. Observations of PSR B0833$-$45 were part of a regular glitch monitoring program, where we regularly monitor 24 pulsars (i) to detect glitches, (ii) model timing noise, and (iii) model post-glitch recovery. This work presents a comprehensive analysis of the Vela pulsar's rotational evolution, spanning from September 2016 to January 2025, as part of our ongoing monitoring campaign \citep{Grover2024, Basu2020}. Our study primarily uses observations from the ORT, supplemented by uGMRT and archival ATNF data to fill temporal gaps and facilitate specialised investigations. 

The ORT \citep{swarup1971ort} is a 530m $\times$ 30m parabolic cylindrical reflector. It is located on a hill with the same slope as the geographic latitude (11\textdegree), allowing it to track sources for about 8 hours and covers a declination range of -60\textdegree \space to +60\textdegree. Pulsar observations at the ORT are carried out at the central observation frequency of 326.5 MHz using the Pulsar Ooty Radio Telescope's New Digital Efficient Receiver, \texttt{PONDER} \citep{Naidu_2015}, which produces real-time coherent dedispersed time series data. The typical cadence of the observations at the ORT is 1--14 days.

The GMRT is an interferometer with 30 parabolic antennas, with a diameter of 45 meters each \citep{Swarup1991GMRT, Ananthakrishnan1995GMRT}. The uGMRT \citep{ugmrt2017} covers a declination range of -53\textdegree to +90\textdegree and supports observations from 150 MHz to 1460 MHz in four bands. The Vela pulsar is observed in uGMRT Band 3 (250--500 MHz) and Band 4 (550--950 MHz). The central square antennas, together with a subset of the nearest arm antennas, were used in a phased array to synthesise a large single dish for observations. The outer arm antennas were not included in the phased array due to significant drift in the phase of the voltage outputs, caused by ionospheric variations, in these antennas. A digital filterbank, implemented in the GMRT correlator, was used to record either 1024 or 2048 channels across the bandpass to obtain frequency-resolved data with a typical sampling time of 655 $\mu$s. The cadence of the observations is 15-30 days. We primarily use frequency-resolved uGMRT data to examine the impact of dispersion measure (DM) offsets on the spin-down rate. Additionally, uGMRT data filled a temporal gap around epoch 60300, providing crucial estimates not covered by ORT due to maintenance downtime. 

The Commonwealth Scientific and Industrial Research Organisation (CSIRO) Data Access Portal is used to access the Parkes pulsar archival observations\footnote{\url{https://data.csiro.au/domain/atnf}} \citep{Hobbs2011} of the Vela pulsar. These data were mainly used to fill temporal gaps in glitch number 2 (MJD 58515), caused by significant observational gaps at ORT. We directly downloaded and used the reduced files (or PSRFITS: \cite{PSRCHIVE_and_PSRFITS2004}) and performed subsequent analysis, as described in the next subsection.

\begin{figure*}
\centering
    \includegraphics[width=0.95\linewidth]{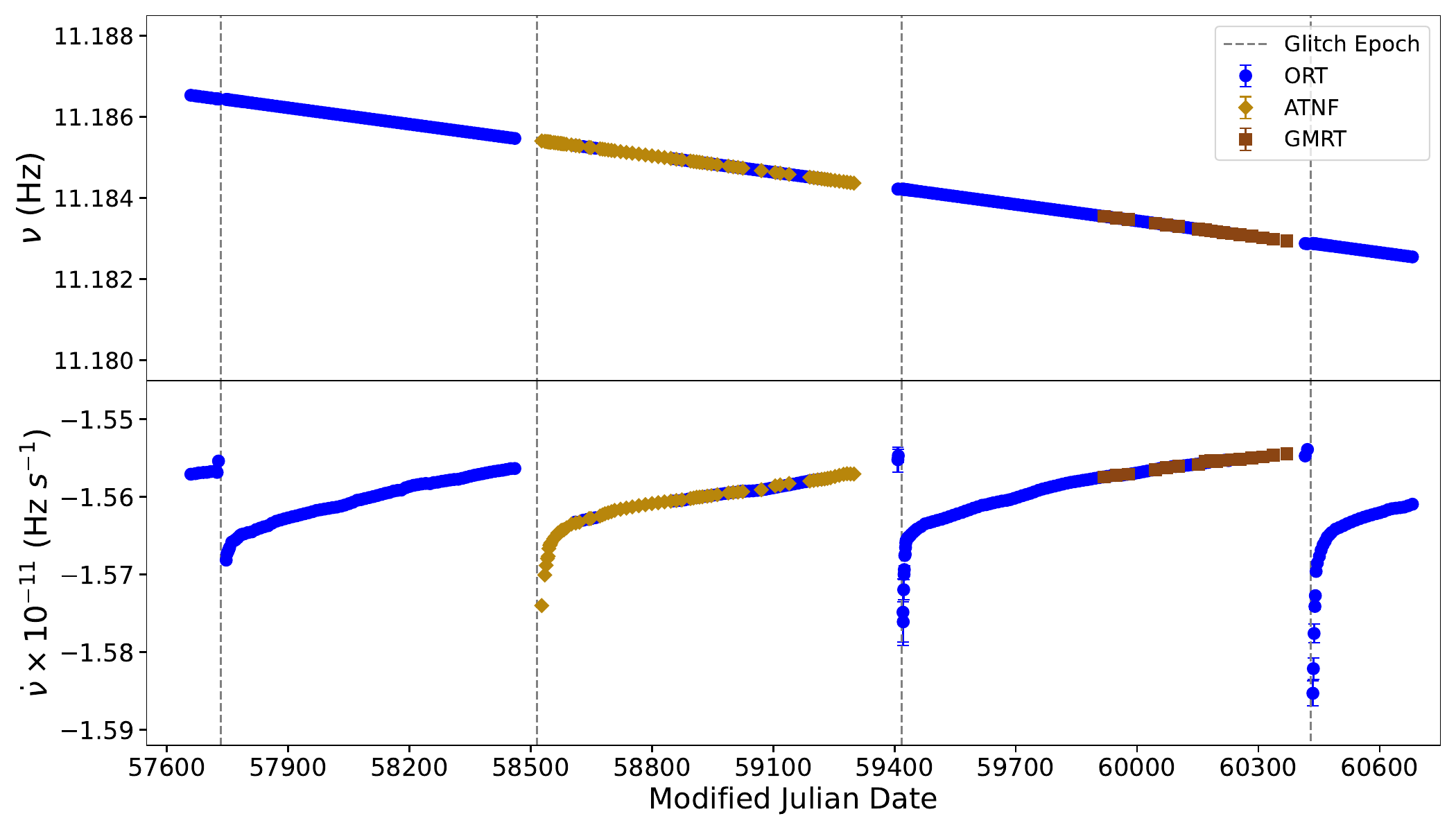}
    \caption{The rotational evolution of the Vela pulsar from September 2016 to January 2025.}
    \label{J0835-4510_all}
\end{figure*}

\subsection{Data Reduction and Timing Analysis}
We used the \texttt{DSPSR} software \citep{DSPSR} and the \texttt{PINTA} pipeline \citep{pinta} to reduce the raw data from the Vela pulsar observed at the ORT and the uGMRT, respectively, into PSRFITS files, using the ephemeris created in our initial observations. The initial analysis, including the generation of a noise-free template and Time of Arrivals (ToAs), was performed using PSRCHIVE \citep{PSRCHIVE_and_PSRFITS2004, PSRCHIVE2012}. The ToAs were obtained through the cross-correlation of the template profile with all other observed profiles using the `pat' command, using the Fourier Phase Gradient (PGS) method \citep{Taylor1992}. The precision timing analysis of the pulsar was done with the help of the pulsar timing software TEMPO2 \citep{Hobbs2006Tempo2, Edwards2006Tempo2}. The DM offsets were estimated using the DMcalc \citep{Krishnakumar2021} script from the frequency-resolved data from uGMRT. The rotational evolution of a pulsar including a glitch can be expressed as a Taylor series expansion \citep{Erbil_2022},

\begin{equation}
\begin{aligned}[c]
    \label{rotationmodel}
        \nu(t)  = & \nu_0 + \dot{\nu}_0(t-t_0) + \frac{1}{2} \Ddot{\nu}_0(t-t_0)^2 + \Delta \nu_g \noindent \\ & + \Delta \dot{\nu}_g(t-t_g) + \Delta \nu_d \exp(-(t-t_g)/\tau_d) \ ,
\end{aligned}
\end{equation}

\noindent where $\nu_0$ is the spin frequency, and $\dot{\nu}_0$ and $\Ddot{\nu}_0$ are the first and second derivatives of the frequency, respectively, at reference epoch $t_0$. $\Delta \nu_g$ and $\Delta \dot{\nu}_g(t-t_g)$ denote the changes in spin frequency and its first derivative at the glitch epoch $t_g$, respectively. The parameters $\Delta \nu_d$ and $\tau_d$ correspond to the amplitude and characteristic decay timescale of the exponential relaxation component following the glitch.

The spin evolution of the Vela pulsar from September 2016 to January 2025, obtained after precision timing, is given in Fig.~\ref{J0835-4510_all}. Each measurement is made with the local timing of several epochs, typically within a 50-200 day data window, contingent on data availability. Exceptions include the first and last few points adjacent to glitches. The resulting measurements are reflected in the precision of error bars, which are smaller than the symbol used in the figure. By choosing multi-epoch windows, we achieve reliable rotational estimates with minimised error bars, providing a robust foundation for studying vortex residuals. This approach effectively averages out temporal effects caused by physical processes, such as ionised interstellar medium scattering-induced broadening, thereby validating the accuracy of measurements aligned with the star's rotation and internal dynamics. These recovery measurements are investigated within the framework of the vortex creep and bending models as described in Section~\ref{vortexCreepModel} and the results are discussed in Section~\ref{ResultsandDiscussions}. 

\subsection{Bayesian Analysis}\label{BayesianAnalysis}
For each recovery, we used Bayesian statistics for model comparison and parameter estimation. Bayesian statistics use Bayes’ theorem to enhance the probability of a hypothesis based on new evidence or data. We used DYNESTY (a dynamic nested sampling package for estimating Bayesian posteriors and evidences; \citep{DYNESTY2020}) to compute the Bayesian evidence and perform model selection, and EMCEE \citep{EMCEE2013} to obtain posterior samples for parameter estimation of the preferred model. For model comparison, the standard \citeauthor{Jeffreys1961} scale is used, which suggests the preferred model based on the strength of evidence \citep{Trotta2008}. Bayesian inference is also used to predict the epoch of the next glitch using data from previous glitches.

The priors are selected based on the initial knowledge of each parameter, as informed by the literature \citep{Erbil_2022, Flanagan1990, Alpar1989}. This knowledge provides reasonable initial ranges that can be further refined or broadened based on the results. For instance, parameters such as the fractional moment of inertia ($I_{\rm e}/I$ or $I_{\rm a}/I$) is generally positive and below unity. The recoupling timescale in exponential regions ($\tau_{\rm e}$) depends on the number of exponential components (three-, two-, or one-component models). Typically, for the Vela pulsar, it has been observed that the first component lies within 0.1–10 days, with the second and third components approximately 10 and 100 times longer, respectively.

For the linear relaxation region, the prior for recoupling timescale ($\tau_{\rm nl}$) is around 10-300 days, and the offset time ($t_0$) is 300 to 1200, with the initial guess often chosen close to the observed time to the next glitch. For the Vortex bending studies, the prior on amplitude, phase, and decay constant were set as 0--1, 0--2$\pi$, and 0--2000 days, respectively. All the priors were considered as uniform priors and listed in Table~\ref{table_priors}. While these priors adequately describe many recoveries, particularly those observed in the Vela pulsar, their applicability may vary depending on the nature of the recovery. As each recovery exhibits unique characteristics, a flexible and adaptive approach to defining priors is essential. We model the data assuming independent Gaussian measurement uncertainties. The corresponding log-likelihood function is given by
\begin{equation}
\ln \mathcal{L}(\boldsymbol{\theta}) = -\frac{1}{2}\sum_{i}
\left[
\frac{\left(d_i - m_i(\boldsymbol{\theta})\right)^2}{\sigma_i^2}
+ \ln\left(2\pi \sigma_i^2\right)
\right],
\end{equation}
where $d_i$ denotes the observed data, $m_i(\boldsymbol{\theta})$ is the model prediction for parameters $\boldsymbol{\theta}$, and $\sigma_i$ represents the corresponding measurement uncertainty.

\begin{table}[h]
\centering
\caption{Prior ranges used for the parameter estimation of each recovery. The left column represents the parameter, and the right column displays the values with their corresponding units. The symbols $I_{\rm e1}/I$, $\tau_{\rm e1}$,$I_{\rm e2}/I$, $\tau_{\rm e2}$,$I_{\rm e3}/I$, $\tau_{\rm e3}$ represent the fractional moment of inertia and the decay timescales for the first, second and third components of exponential recovery, respectively. $I_{\rm a}/I$, $\tau_{\rm nl}$, $t_0$ denote the fractional moment of inertia, the decay timescales, and offset time associated with the linear relaxation, respectively. And $A$ is the amplitude, $\phi$ is the phase, and $\tau$ is the exponential decay constant for oscillations.}
\renewcommand\arraystretch{1.5}
\setlength\tabcolsep{6pt}
\begin{tabular}{|cc||cc|}
\hline
Parameters & Values & Parameters & Values \\ 
\hline
\centering
$I_{\rm e1}/I$ & 0--1 & $I_{\rm a}/I$ & 0--1\\
$I_{\rm e2}/I$ & 0--1 & $\tau_{\rm nl}$ &  10--300 days \\ 
$I_{\rm e3}/I$ & 0--1 & $t_0$ &  300--1200 days  \\
$\tau_{\rm e1}$ &  0.1--10 days & $A$ &  0--1  \\
$\tau_{\rm e2}$ &  5--100 days & $\phi$ & 0--2$\pi$ \\
$\tau_{\rm e3}$ &  10--400 days & $\tau$ & 0--2000 days \\
\hline
\end{tabular}
\label{table_priors}
\end{table}

The post-glitch recoveries are studied within the framework of the vortex creep model and the vortex bending model \citep{Erbil_2023}, discussed in the previous section. A quick picture of the analysis process is as follows: We first model the frequency derivative measurements from the data using the vortex creep model. Then, we estimate the residuals obtained by subtracting the predicted values from the vortex creep model from the measured values. These residuals were investigated within the framework of the vortex bending model. The generalised Lomb-Scargle periodogram technique \citep{Lomb1976, Scargle1982, VanderPlas2018} was employed to detect periodicities in the residuals. We used the {\tt Astropy}~\citep{Astropy2013} implementation of the generalised Lomb-Scargle periodogram.
The  Peaks with false alarm probabilities (FAP) below 0.1\% or Z-score above 3$\sigma$ were considered significant for vortex bending studies. FAP estimation was performed using the~\citet{Baluev2008} method. Alternative methods, including approximations based on effective independent frequencies and bootstrap resampling, yielded similar results. 
The significance of each detected peak is quantified using the Lomb–Scargle false-alarm probability (FAP). For interpretive purposes, we also report the equivalent z-score values~\citep{Cowan2011, Ganguly_Desai_2017}. The results of our analysis are presented in the next section.

\section{Results and Discussions}\label{ResultsandDiscussions}
Our pulsar monitoring program with the ORT and the uGMRT started in 2015. The previous results were reported in \cite{Basu2020} and \cite{Grover2024}. In this work, we present an analysis of a new Vela glitch at MJD 60429.9. Further, we demonstrate the spin-down rate evolution for the Vela pulsar since the beginning of the monitoring program.  Our primary focus is to provide the theoretical interpretation of glitches using the post-glitch recovery phase and to obtain measurements 
for the re-coupling timescales and fractional moment of inertia for various layers that participate in a glitch. We have estimated the theoretical time to the next glitch through the vortex creep model and used Bayesian inference to predict the time for the next glitch. All the predictions match well with the observed inter-glitch times.

\subsection{Glitches}
Here, we summarise the last three glitches in the Vela pulsar: MJD 57734.4, 58517, and 59417.6, detected in our monitoring program \citep{Grover2024, Basu2020}. The recoveries for each of these glitches are studied within the framework of the vortex creep model to investigate the structure of the Vela pulsar. Furthermore, we report the detection and analysis of the recent large glitch in the Vela pulsar in 2024. The characteristics and parameters of glitches are listed in Table~\ref{table_glitch_paramters}. 

\begin{table*}[ht]
\centering
\caption{Glitches in the Vela pulsar detected in our monitoring program. The columns from left to right represent the glitch number, glitch epoch, pre-glitch spin frequency, pre-glitch spin-down rate, fractional change in spin frequency (glitch magnitude), and fractional change in the spin-down rate and the reference.}
\renewcommand\arraystretch{2}
\setlength\tabcolsep{6pt}
\begin{tabular}{|c|c|cc|cc|c|}
\hline
Number & Epoch (MJD) & Pre-glitch $\nu$ (Hz) & Pre-glitch $\Dot{\nu}$ (Hz s$^{-1}$) & $\frac{\Delta \nu}{\nu}$ ($\times 10^{-9}$) & $\frac{\Delta \Dot{\nu}}{\Dot{\nu}}$ $\times 10^{-3}$ & Reference\\
\hline
\centering
G1 & 57734.4(2) & 11.186433252(6) & -1.556383(8)$\times 10^{-11}$ & 1433.2(9) & 5.595(9) & \cite{Basu2020}\\
G2 & 58517(7)  & 11.185397242(3) & --1.555659(4)$\times 10^{-11}$ 
 & 2471(6) & 6(2) & \cite{Grover2024} \\
G3 & 59417.6(1)  &  11.184208478(5) & --1.5550(8)$\times 10^{-11}$ & 1235(5) & 8.0(7) & \cite{Grover2024} \\
G4 & 60429.9(1)  & 11.182859362(3) & --1.5530(8)$\times 10^{-11}$ & 2396(2) & 23(1) & This work \\
\hline
\end{tabular}
\label{table_glitch_paramters}
\end{table*}

\begin{figure}
\centering
\includegraphics[width=\linewidth]{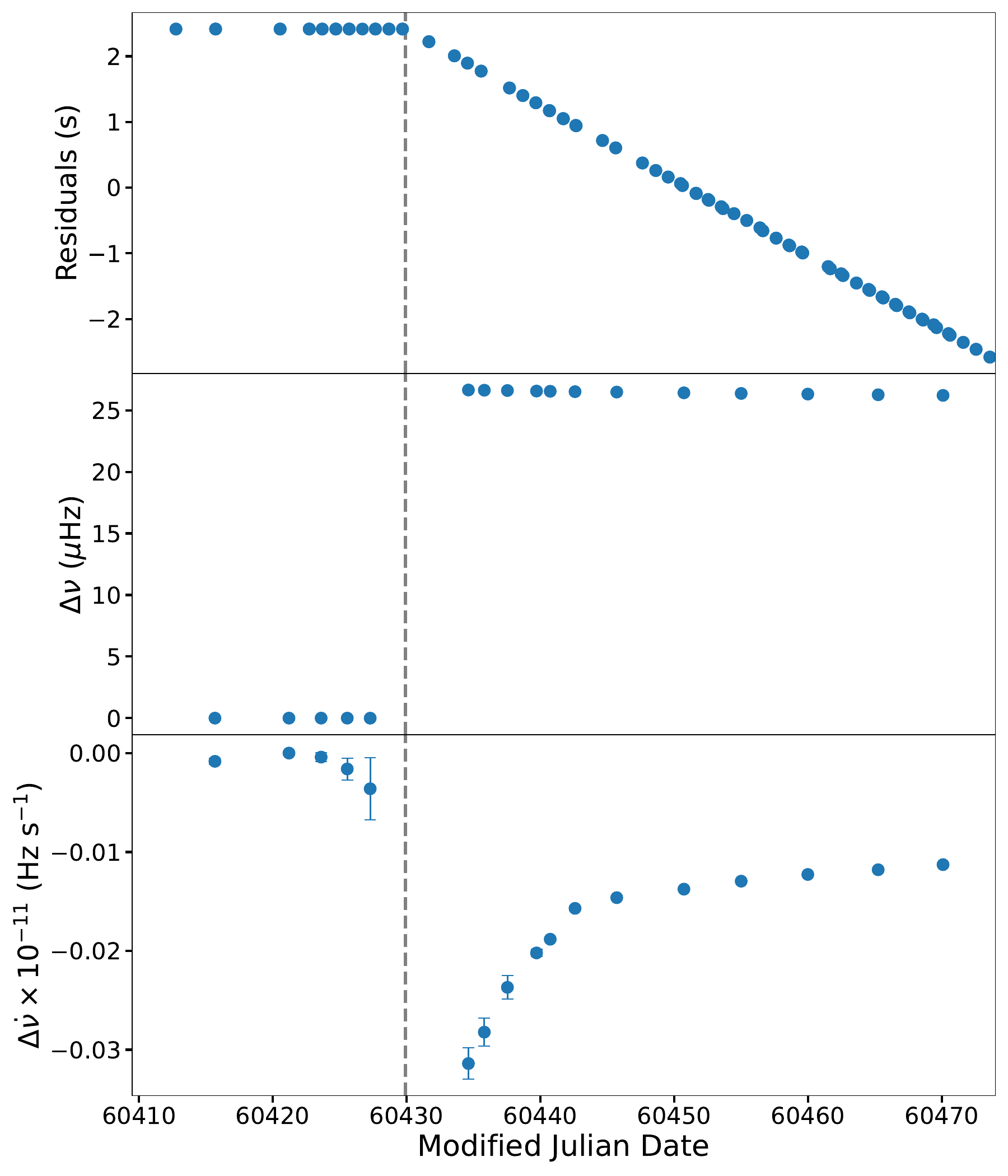}
\caption{Glitch observed in PSR J0835--4510 on MJD 60429.9. The top panel represents the timing residuals. The middle and bottom panels display the evolution of $\Delta \nu$ and $\Delta \dot\nu$, respectively. The vertical dashed line indicates the glitch epoch.}
\label{J0835_glitch}
\end{figure}

We present the parameters of the Vela 2024 glitch, which occurred around MJD 60429.9. This glitch has been reported by several groups \citep{Zubieta2024ATel16608_Vela, Campbell-Wilson2024ATel16610_Vela, Grover2024ATel16611_Vela, Palfreyman2024ATel16615_Vela, Wang2024ATel16619_Vela, Zubieta2025_Vela}.  Here, we present updates in our preliminary results reported in \cite{Grover2024ATel16611_Vela}. The glitch is estimated to occur at MJD 60429.9(1), with the fractional increase in the spin frequency, and its time derivative is calculated as $2396(2)\times10^{-9}$ and $23(1)\times10^{-3}$, respectively. The results of the glitch epoch and spin parameters agree with the reported estimates of other monitoring programs. A more precise measurement of glitch epoch as 60429.86962(4) is also presented in \cite{Palfreyman2024ATel16615_Vela} with higher cadence observations. The timing residuals, the evolution of the spin frequency, and the spin-down rate are given in Fig.~\ref{J0835_glitch}. The post-glitch recovery for this glitch is also investigated within the framework of the vortex creep model in the next subsection. 

\subsection{Post-Glitch Recoveries}
In this section, we present the post-glitch recovery behaviour of the Vela pulsar. The spin-down rate has been analysed from MJD 57651 to 60700. During this period, the Vela pulsar has displayed four glitches. The post-glitch recovery of the Vela pulsar is regular, i.e., it follows similar behaviour. The recovery for this pulsar usually consists of exponential and linear relaxations well described by Model 3 of Section \ref{vortexCreepModel}. We have investigated all post-glitch recoveries for this pulsar within the framework of the vortex creep model for the glitches listed in Table~\ref{table_glitch_paramters}, and the resulting fit parameters obtained using the vortex creep model are listed in Table~\ref{table_post-glitch}. Now, we concisely describe the post-glitch relaxation(s) for individual cases.

\begin{table*}[h]
\centering
\caption{The parameters estimated by fitting the vortex creep model to the observed post-glitch recovery data for the last four glitches in the Vela pulsar. The first column represents the glitch epoch. Columns 2-7 represent the fractional moment of inertia, and corresponding decay timescales with 95\% credible intervals for the first, second, and third exponential recovery components (Equation~\ref{eqn_exp_recovery}), respectively. Columns 8 and 9 provide the fractional moment of inertia for linear recovery and the recoupling timescale with 95\% credible intervals, respectively. The offset time with a 95\% credible interval is given in column 10 (Equation~\ref{eqn_linear_recovery}). The observed inter-glitch interval is given in column 11, and column 12 contains the predicted inter-glitch time using MCMC sampling with a 68\% credible interval.}
\renewcommand\arraystretch{2}
\setlength\tabcolsep{6pt}
\begin{tabular}{|c|cccccc|ccc|c|c|}
\hline
Epoch & $I_{e1}/I$ & $\tau_{e1}$ & $I_{e2}/I$ & $\tau_{e2}$ & $I_{e3}/I$ & $\tau_{e3}$ & $I_a/I$ & $\tau_{nl}$ & $t_0$ & $t_{obs}$ & $t_{B}$\\
(MJD) & ($\times 10^{-3}$) & (days) & ($\times 10^{-3}$) & (days) & ($\times 10^{-3}$) & (days) & ($\times 10^{-3}$) & (days) & (days) & (days) & (days) \\
\hline
\centering
57734.4(2) & $4.5^{+2.4}_{-1.3}$ & $4.9^{+1.1}_{-1.0}$ & $3.9^{+0.3}_{-0.3}$ & $29.3^{+1.7}_{-1.6}$ & $58.4^{+2.8}_{-2.7}$ & $298.5^{+8.8}_{-8.8}$ & $4.4^{+0.1}_{-0.1}$ & $62.2^{+2.4}_{-2.4}$ & $774.0^{+2.2}_{-2.3}$ & 783 & [787, 1083] \\
58517(7) & --- & --- & $9.8^{+0.3}_{-0.3}$ & $7.4^{+0.2}_{-0.2}$ & $10.4^{+0.1}_{-0.1}$ & $59.5^{+0.4}_{-0.4}$ & $4.2^{+0.0}_{-0.0}$ & $22.4^{+1.5}_{-1.6}$ & $921.6^{+0.4}_{-0.4}$ & 902 & [752, 1033] \\
59417.6(1) & $9.4^{+5.7}_{-3.7}$ & $1.6^{+0.4}_{-0.3}$ & $3.4^{+0.2}_{-0.2}$ & $13.8^{+0.7}_{-0.7}$ & $19.7^{+2.7}_{-2.6}$ & $228.0^{+9.2}_{-9.8}$ & $9.1^{+0.2}_{-0.2}$ & $212.8^{+1.1}_{-1.1}$ & $469.1^{+5.5}_{-5.6}$ & 1012 & [855, 1123] \\
60429.9(1) & $7.0^{+2.0}_{-2.0}$ & $2.4^{+0.4}_{-0.4}$ & $5.3^{+0.3}_{-0.3}$ & $12.1^{+0.5}_{-0.4}$ & $20.5^{+0.9}_{-1.4}$ & $135.1^{+4.7}_{-5.9}$ & $6.0^{+0.3}_{-0.4}$ & $282.0^{+17.3}_{-67.9}$ & $1182.8^{+16.6}_{-69.1}$ & --- & [820, 1076]\\
\hline
\end{tabular}
\label{table_post-glitch}
\end{table*}

\subsubsection{Glitch No. 1 -- MJD 57734.4(2)}

\begin{figure}
\centering
    \includegraphics[width=\linewidth]{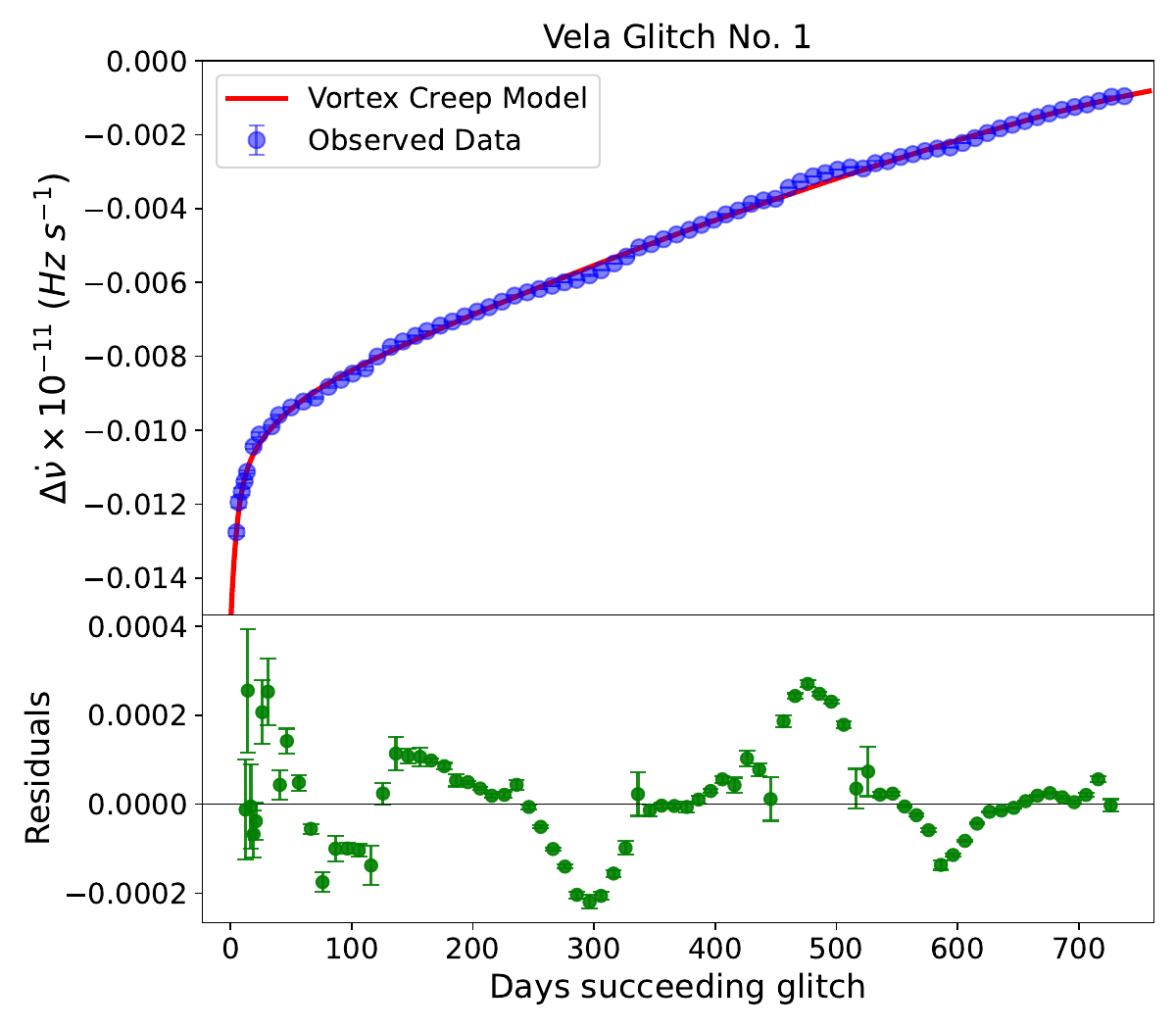}
    \caption{The post-glitch recovery of the Vela glitch G1. The top panel shows the observed change in the spin-down rate relative to the pre-glitch spin-down rate (blue dots) and the best-fit vortex creep model predictions (red solid curve). The bottom panel displays the residuals obtained by subtracting the best-fit values from the measurements of the spin-down rate on the scale of $10^{-11}$ Hz s$^{-1}$.}
    \label{J0835-4510_57734_recovery}
\end{figure}

G1, or glitch at MJD 57734.4(2) \citep{Sarkissian2017, Palfreyman2018_Vela, Ashton2019, Basu2020, Erbil_2020, Montoli2020, Dunn2025}, was detected with a fractional change in the rotational frequency of 1433(1)$\times 10^{-9}$, and a fractional change in the spin-down rate as 5.59(1)$\times 10^{-3}$. We present the post-glitch relaxation modelling for this glitch in Fig.~\ref{J0835-4510_57734_recovery}. For this recovery, the Bayesian model comparison prefers the presence of three-exponential + linear recovery (Model 3c) over all other models. The estimated median values for parameters are: the fractional moment of inertia for the first, second, and third exponential components are $4.5 \times 10^{-3}$, $3.9 \times 10^{-3}$ and $58.4 \times 10^{-3}$ respectively, with the decay timescales of $4.9$, $29.3$ and $298.5$ days, respectively. The median fractional moment of inertia for the non-linear creep region is $4.4 \times 10^{-3}$ with a recoupling timescale of 62 days. The offset time is measured as $774.0$ days, close to the observed inter-glitch time of 783 days. The observed time is consistent with the vortex creep model's predicted range and closely matches the Bayesian prediction, which estimates a 68\% credible interval of 787 to 1083 days. The fit parameters, along with their 95\% credible intervals, are listed in Table~\ref{table_post-glitch}. 

\begin{figure}
\centering
    \includegraphics[width=\linewidth]{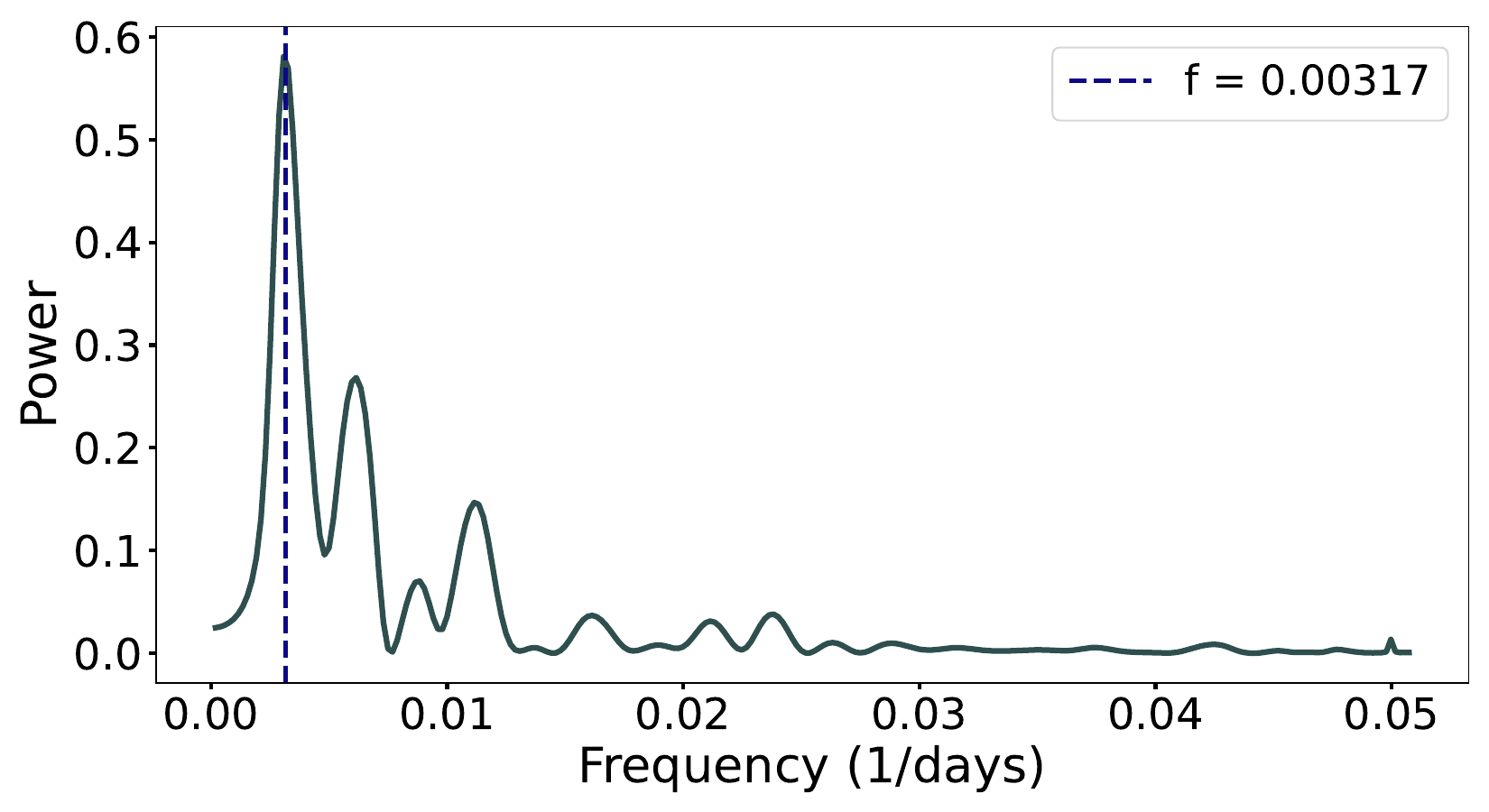}
    \caption{The Lomb-Scargle Periodogram for the vortex residuals of G1 in the Vela pulsar.}
    \label{J0835-4510_57734_lsp}
\end{figure}

The bottom part of Fig.~\ref{J0835-4510_57734_recovery} shows the residuals obtained after subtracting the vortex creep model predictions from the observed data and is in units of $10^{-11}$ Hz s$^{-1}$. These vortex residuals, spanning from 200 days onward, were used for vortex bending studies, excluding the initial 200 days due to the significant impact of exponential recoveries. The generalised Lomb–Scargle periodogram (Fig.~\ref{J0835-4510_57734_lsp}) reveals a prominent peak corresponding to a period of 315(1) days using the least-squares interpretation of the periodogram \citep{VanderPlas2018}. This signal exhibits strong significance, with a false alarm probability (FAP) of $2.3 \times 10^{-7}$ (Z-score: 5.0$\sigma$). A secondary peak at approximately 165 days has a FAP of 0.14, while all remaining peaks yield FAP values of 1, indicating low significance.

\begin{figure}
\centering
    \includegraphics[width=\linewidth]{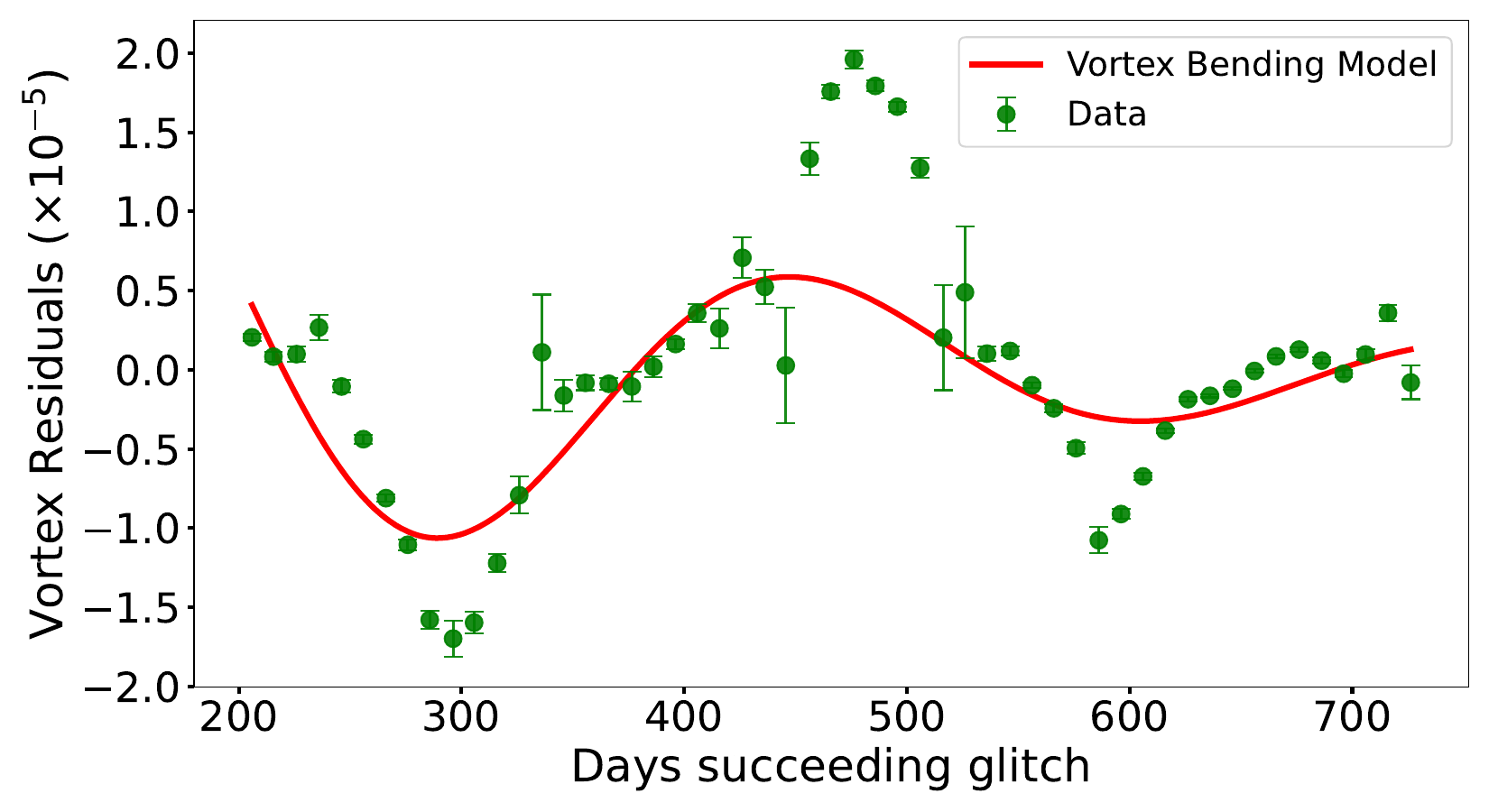}
    \caption{The vortex residuals of G1 and the vortex bending model. The vortex residuals are in units of $10^{-11}$ Hz s$^{-1}$.}
    \label{J0835-4510_57734_residuals}
\end{figure}

Fixing the period at 315(1) days, we investigated the vortex bending hypothesis and found that the damping model, comprising a sinusoidal term and a damping factor (Hypothesis II), was strongly favoured over Hypotheses I and III. Additionally, we examined whether the observed peak resulted from shocks caused by the current glitch or the previous one, concluding that the oscillations decisively originated from the current glitch. The comparison of vortex residuals and vortex bending model is shown in Fig.~\ref{J0835-4510_57734_residuals} and the estimated parameters for the damped model (Hypothesis II) are: Amplitude = $3.2_{-0.2}^{+0.2} \times 10^{-5}$, Phase = $5.04_{-0.02}^{+0.02}$, and $\tau = 266_{-8}^{+8}$ days. Note that vortex residuals are in scale of $10^{-11}$ Hz s$^{-1}$, so the effective amplitude is $3.2_{-0.2}^{+0.2} \times 10^{-16}$. 

The vortex residuals fit with the vortex bending model well for around 200-400 days post-glitch duration, but deviate thereafter. Including a second peak improves the fit, but the generalised Lomb-Scargle periodogram indicates a FAP of 14\% for this peak using the Baluev method (with a minimum FAP of 5.3\% obtained through approximations based on effective independent frequencies). Given that the FAP of the second peak is greater than 0.1\%, we exclude the second peak from our analysis. The other peaks with a high FAP are likely to originate from preceding glitches, and their effect on the present glitch recovery is insignificant. The post-glitch recovery modeling for this pulsar is also presented by \cite{Erbil_2022} using the Fermi-LAT observations. However, they only detected one exponential component. We are reporting three exponential recoveries made possible by much higher cadence observations, which allow us to investigate the vortex residuals within the framework of the vortex bending model. 

\subsubsection{Glitch No. 2 -- MJD 58517(7)}
\begin{figure}
\centering
    \includegraphics[width=\linewidth]{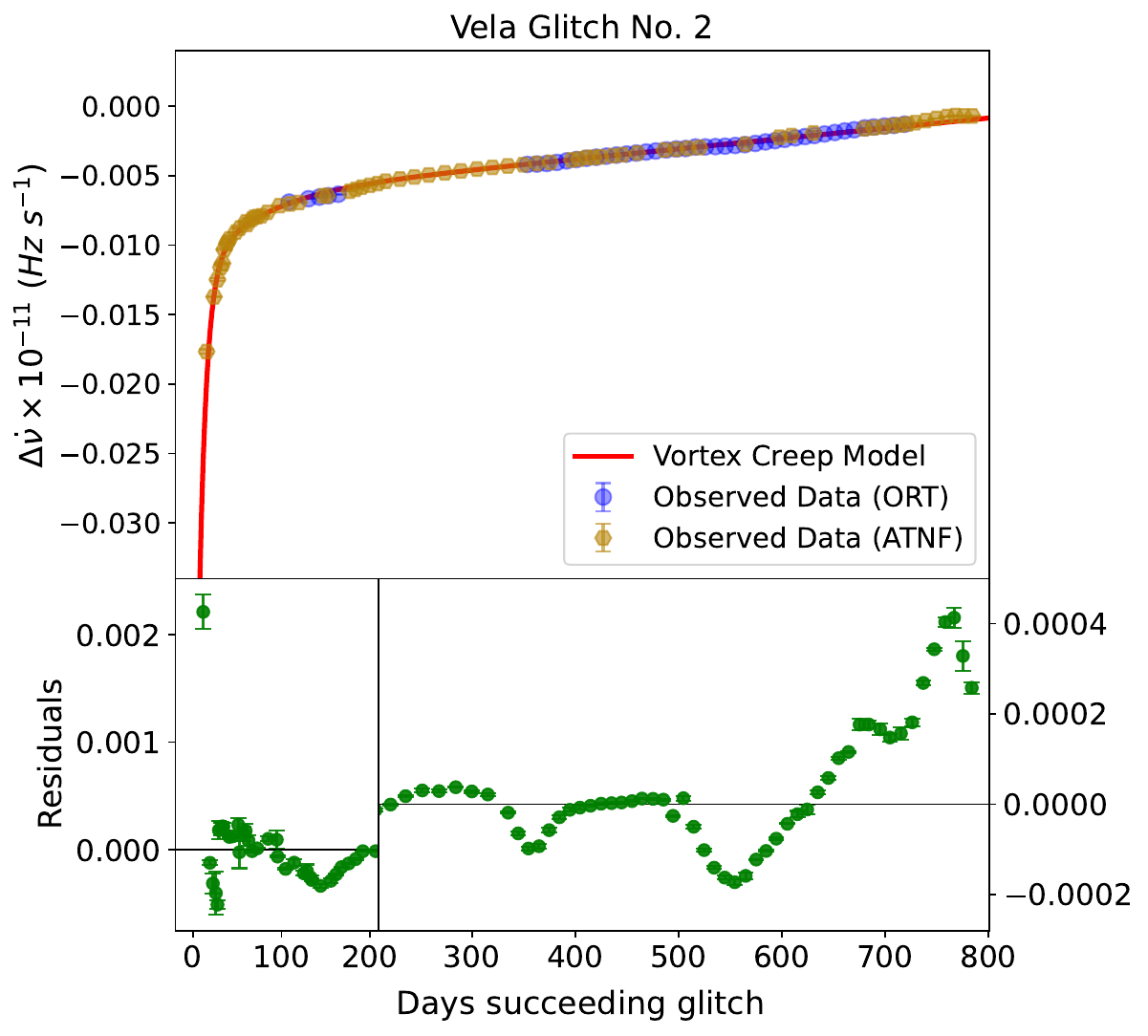}
    \caption{The post-glitch recovery of the Vela glitch G2. The top panel shows the observed change in the spin-down rate relative to the pre-glitch spin-down rate (blue dots our data, orange dots the ATNF/Parkes public data) and the best-fit vortex creep model predictions(red solid curve). The bottom panel displays the residuals obtained by subtracting the best-fit values from the measurements of the spin-down rate on the scale of $10^{-11}$ Hz s$^{-1}$.}
    \label{J0835-4510_58515_recovery}
\end{figure}

The glitch at MJD 58517(7), G2 in the Vela pulsar, \citep{Sarkissian2019ATel12466, Kerr2019ATel12481, LopezArmengol2019ATel12482, lower2020_utmost2, Erbil_2022, Grover2024, 
Dunn2025}, was observed with a fractional change in the rotational frequency of 2471(6)$\times 10^{-9}$, and a fractional change in the spin-down rate as 6(2)$\times 10^{-3}$. Our analysis focuses on modelling the post-glitch relaxation behaviour, given in Fig.~\ref{J0835-4510_58515_recovery}. Our ORT and uGMRT observations weren't conducted during the initial exponential relaxation phase of this glitch, and we detected only non-linear creep recovery. To proceed with our post-glitch analysis, we supplemented our dataset with publicly available ATNF pulsar data to fill temporal gaps. Additionally, we used estimates derived from analysis using data closer to the glitch from \cite{lower2020_utmost2}, specifically the glitch epoch (58515.6) and the fractional change in rotational frequency and its derivative (2501 and 8.7, respectively), for our recovery studies.

The Bayesian model comparison reveals that the model with two exponential terms and a linear recovery term (Model 3b) provides the best description of this recovery. The measured median values for the recovery parameters are as follows: the fractional moment of inertia for the second and third exponential components are $9.8 \times 10^{-3}$ and $10.4\times 10^{-3}$ respectively, with the decay timescales of $7.4$ and $59.5$ days respectively. The fractional moment of inertia for the non-linear creep region is $4.2 \times 10^{-3}$ with a recoupling timescale of $22.4$ days. The offset time is estimated as $921.6$ days. The observed inter-glitch time of 902 days is consistent with both the predicted range of the vortex creep model and Bayesian statistics. The median value predicted by Bayesian statistics is 893 days, very close to the observed time. These parameters are also summarised in Table~\ref{table_post-glitch}. Unlike other glitches that were observed with very high cadence using the ORT, this glitch was not monitored with such precision, and therefore, the vortex-bending oscillations are not well observed and studied in this work. Our post-glitch recovery modelling results differ from those of \cite{Erbil_2022}, who reported a single exponential component using Fermi-LAT observations. Our analysis benefits from a higher cadence and reveals two well-constrained, distinct exponential recoveries.

\subsubsection{Glitch No. 3 -- MJD 59417.6(1)}
\begin{figure}
\centering
    \includegraphics[width=\linewidth]{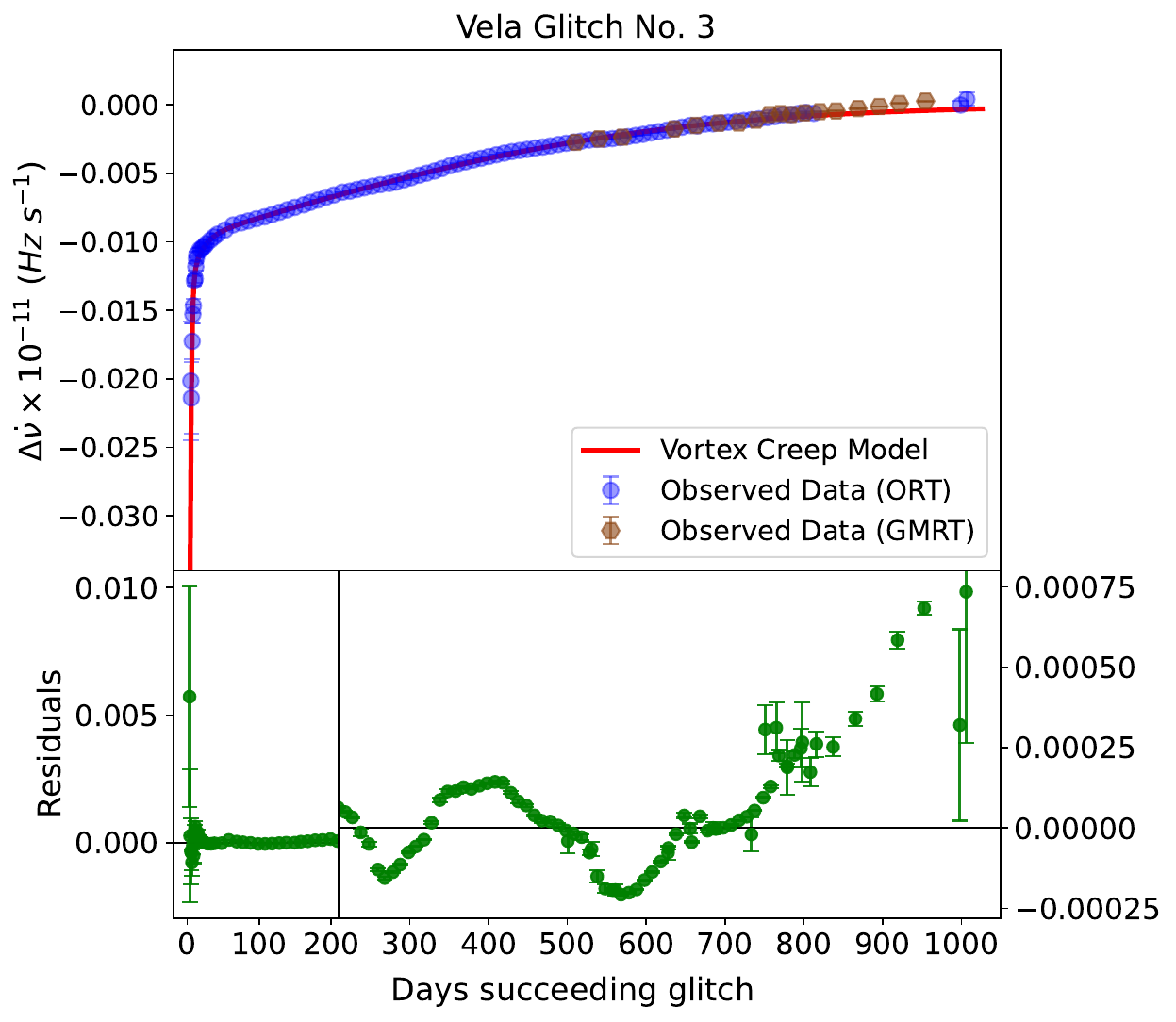}
    \caption{The post-glitch recovery of the Vela glitch G3. The two diagrams are similar to Fig.~\ref{J0835-4510_57734_recovery} and~\ref{J0835-4510_58515_recovery}.}
    \label{J0835-4510_59417_recovery}
\end{figure}

G3, or the glitch at MJD 59417.6(1) \citep{Sosa-Fiscella2021ATel14806, Dunn2021ATel14807, Olney2021ATel14808, Singha2021ATel14812, Zubieta_2023, Grover2024, Dunn2025} was observed to have a fractional spin up in frequency $1235(5) \times 10^{-9}$ and a fractional spin up in frequency-time derivative $8.0(7) \times 10^{-3}$. We present the post-glitch relaxation behaviour for this glitch in Fig.~\ref{J0835-4510_59417_recovery}. In addition to the ORT, we leverage observed data from the uGMRT as well, with two motivations: (i) to confirm that the residuals stem from the neutron star's interior, rather than the ionised interstellar medium by using DM offsets in the uGMRT analysis and (ii) to fill the temporal gap between 850 to 1000 days subsequent to the glitch date. 

High-cadence ORT observations of this glitch enable us to model three exponential components present in the recovery; Model 3c emerged as the top choice in Bayesian model comparisons, outperforming all other models. The calculated median values for the parameters are as follows: the fractional moment of inertia for the first, second and third exponential components are $9.4 \times 10^{-3}$ and $3.4 \times 10^{-3}$ and $19.7 \times 10^{-3}$ respectively, with the decay timescales of $1.6$, $13.8$ and $228.0$ days, respectively. The fractional moment of inertia for the non-linear creep region is $9.1 \times 10^{-3}$ with a recoupling timescale of $212.8$ days. The offset time $t_0$ is estimated as $469.1$ days, which deviates significantly from the observed inter-glitch time primarily due to the large value of $\tau_{nl}$ (equivalent to $\tau_{e3}$). However, the vortex creep model predicted time for the next glitch lies between $t_0-3\tau_{nl}$ to $t_0+3\tau_{nl}$, which aligns well with the observed time. The median Bayesian prediction of 989 days is near the observed interval of 1012 days, falling well within the 68\% credible interval. These measured parameters are listed in Table~\ref{table_post-glitch}.

The relaxation behaviour of this glitch has never been investigated before. The lower panel of Fig.~\ref{J0835-4510_59417_recovery} shows the vortex residuals, obtained after subtracting the vortex creep model predictions from the observed data, in scale of $10^{-11}$ Hz s$^{-1}$. We analysed residuals between 200 and 700 days for vortex bending studies, excluding the initial 200 days due to dominant exponential recovery effects, and data beyond 700 days requires further investigation in existing theory. 

\begin{figure}
\centering
    \includegraphics[width=\linewidth]{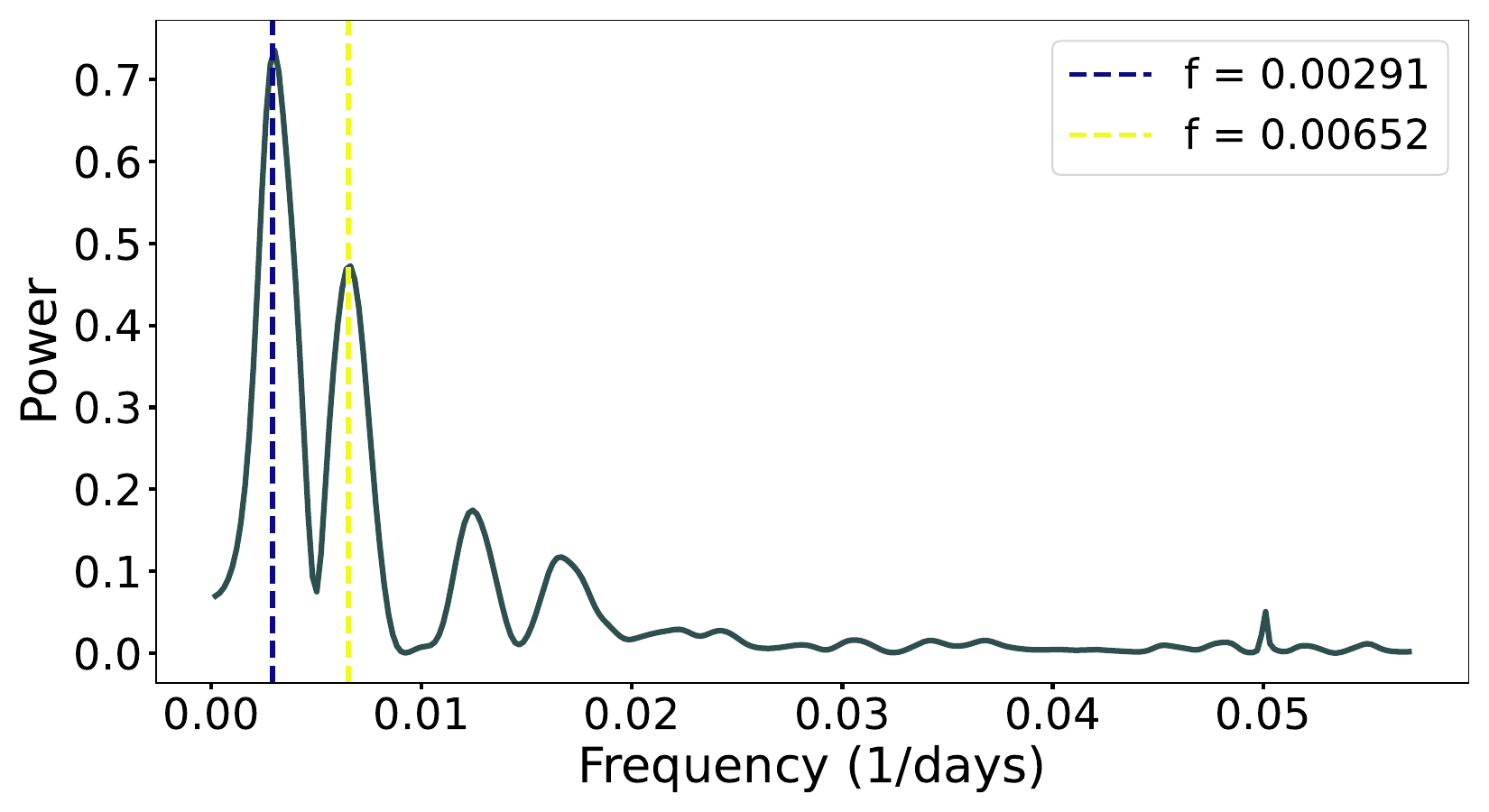}
    \caption{The Lomb-Scargle Periodogram for the vortex residuals of G3 in the Vela pulsar.}
    \label{J0835-4510_59417_lsp}
\end{figure}

\begin{figure}
\centering
    \includegraphics[width=\linewidth]{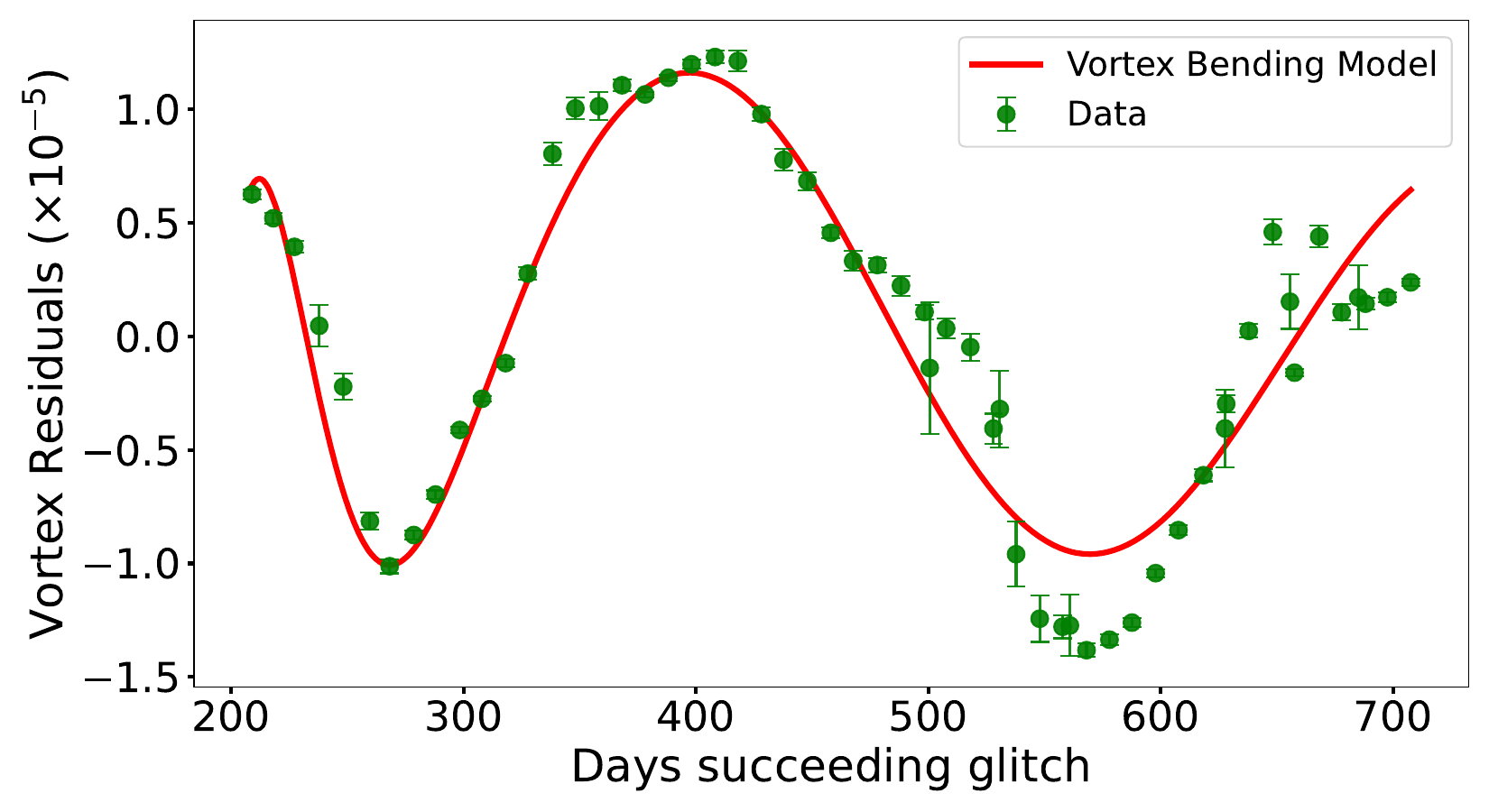}
    \caption{The vortex residuals of G3 and the vortex bending model. The vortex residuals are in units of $10^{-11}$ Hz s$^{-1}$.}
    \label{J0835-4510_59417_residuals}
\end{figure}

The Lomb–Scargle periodogram analysis of the vortex residuals (Fig.~\ref{J0835-4510_59417_lsp}) reveals several peaks, but only the first two are statistically significant. The strongest peak yields a FAP of 4.6 $\times$ 10$^{-13}$ and a Z-score of 7.2$\sigma$, while the latter peak has a FAP of 2.46 $\times$ 10$^{-5}$ and a Z-score of 4.1$\sigma$. The third peak, with a period of around 81 days, has a FAP of 0.88 and is therefore not significant; all remaining peaks have FAP values consistent with unity. The corresponding periods of the two significant peaks are approximately 344.0(7) and 153.5(7) days, estimated using the least-squares interpretation of the periodogram \citep{VanderPlas2018}.

These two peaks are used to investigate the vortex residuals within the framework of the vortex bending hypothesis, presented in Fig.~\ref{J0835-4510_59417_residuals}. We observed that Hypothesis II was strongly favoured over Hypotheses I and III. Additionally, we examined whether the observed peak resulted from shocks caused by the current glitch or the previous one, concluding that the oscillations decisively originated from the current glitch. The estimated parameters for the damped model (Hypothesis II) are: (i) Amplitudes, A1 = $1.79_{-0.02}^{+0.02} \times 10^{-5}$, and A2 =$3.39_{-0.79}^{+1.03} \times 10^{-2}$, (ii) Phase, $\phi_1 = 0.52_{-0.01}^{+0.01}$ and $\phi_1 = 4.72_{-0.03}^{+0.03}$, (iii) $\tau_1 = 918.19_{-33.33}^{+34.32}$ and $\tau_2 = 29.77_{-0.99}^{+1.06}$ days. Note that vortex residuals are in scale of $10^{-11}$ Hz s$^{-1}$, so the effective amplitudes are $1.79_{-0.02}^{+0.02} \times 10^{-16}$, and $3.39_{-0.79}^{+1.03} \times 10^{-13}$, respectively. 

\subsubsection{Glitch No. 4 -- MJD 60429.9(1)}
\begin{figure}
\centering
    \includegraphics[width=0.99\linewidth]{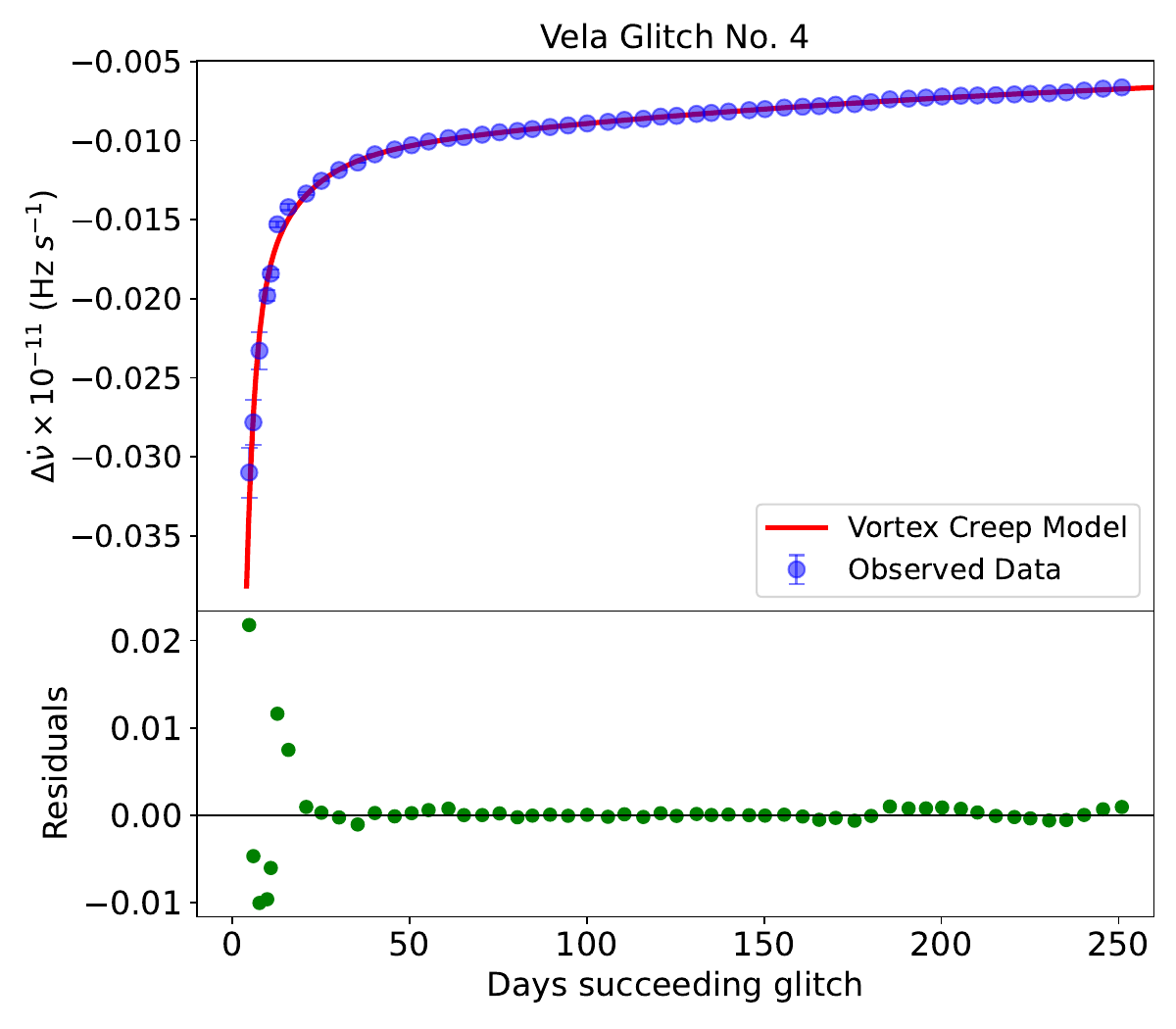}
    \caption{The post-glitch recovery of the Vela glitch G4. The two diagrams are similar to Fig. 
    \ref{J0835-4510_57734_recovery}. }
    \label{J0835-4510_60429_recovery}
\end{figure}

We present the post-glitch relaxation for the recent glitch at MJD 60429.9(2), G4 \citep{Zubieta2024ATel16608_Vela, Campbell-Wilson2024ATel16610_Vela, Grover2024ATel16611_Vela, Palfreyman2024ATel16615_Vela, Wang2024ATel16619_Vela, Zubieta2025_Vela}. The glitch is illustrated in Fig.~\ref{J0835_glitch}, and the comparison of the observed data with the vortex creep model fitting is presented in Fig.~\ref{J0835-4510_60429_recovery}. 

The Bayesian model comparisons selected Model 3c as the best model over all other models. The calculated median values for the parameters are as follows: the fractional moment of inertia for the first, second and third exponential components are $7.0 \times 10^{-3}$ and $5.3 \times 10^{-3}$ and $20.5 \times 10^{-3}$ respectively, with the decay timescales of $2.4$, $12.1$ and $135.1$ days, respectively. The median fractional moment of inertia for the non-linear creep region is $6.0 \times 10^{-3}$ with a recoupling timescale of $282.0$ days. The median value for the offset time is $1182.8$ days. These parameters, along with their 95\% credible intervals, are listed in Table~\ref{table_post-glitch}. The parameters for the linear creep regions (or exponential recovery) are well-constrained. However, those dependent on the non-linear creep region will likely be refined in the future as more data becomes available. The Bayesian analysis forecasts a median time for the next glitch at MJD 61377.7, with a 68\% credible interval spanning from MJD 61249 to 61506.

\subsection{Correlations and Braking Index}\label{Braking index}
\begin{figure}
    \centering
    \includegraphics[width=\linewidth]{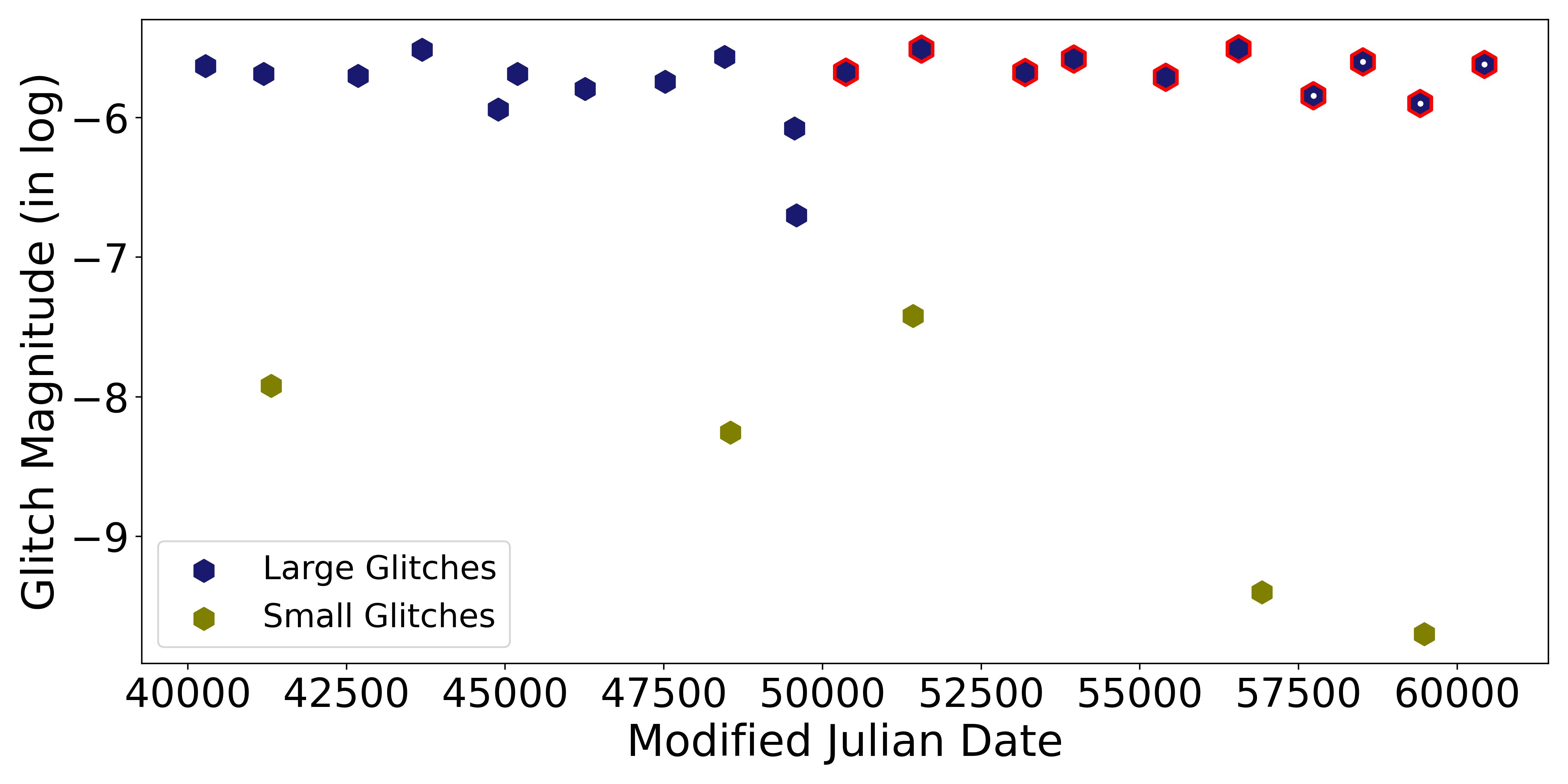}
    \caption{Glitch magnitude vs the detection epoch (in MJD). Large glitches are displayed in midnight blue, while small ones are shown in olive. The alternating pattern of increasing and decreasing magnitudes among the large glitches is highlighted with a red boundary. The last four glitches, marked with small white dots, represent the events studied in this work.}
    \label{fig: GAvsMJD}
\end{figure}

To date, 26 glitches have been reported in the Vela pulsar\footnote{\url{http://www.jb.man.ac.uk/pulsar/glitches.html}} \citep{Espinoza2011, Basu2022}, with 21 of them having large magnitudes (> 10$^{-7}$). Fig.~\ref{fig: GAvsMJD} presents these 26 events. A pattern is observed among the large glitches, where their magnitudes exhibit an alternating trend: a glitch with a lower magnitude is typically followed by one with a higher magnitude, and vice versa. This alternating behaviour is particularly evident in the last 10 glitches (i.e., after MJD 50000), as highlighted by the red boundary in Fig.~\ref{fig: GAvsMJD}.

\begin{figure}
    \centering
    \includegraphics[width=\linewidth]{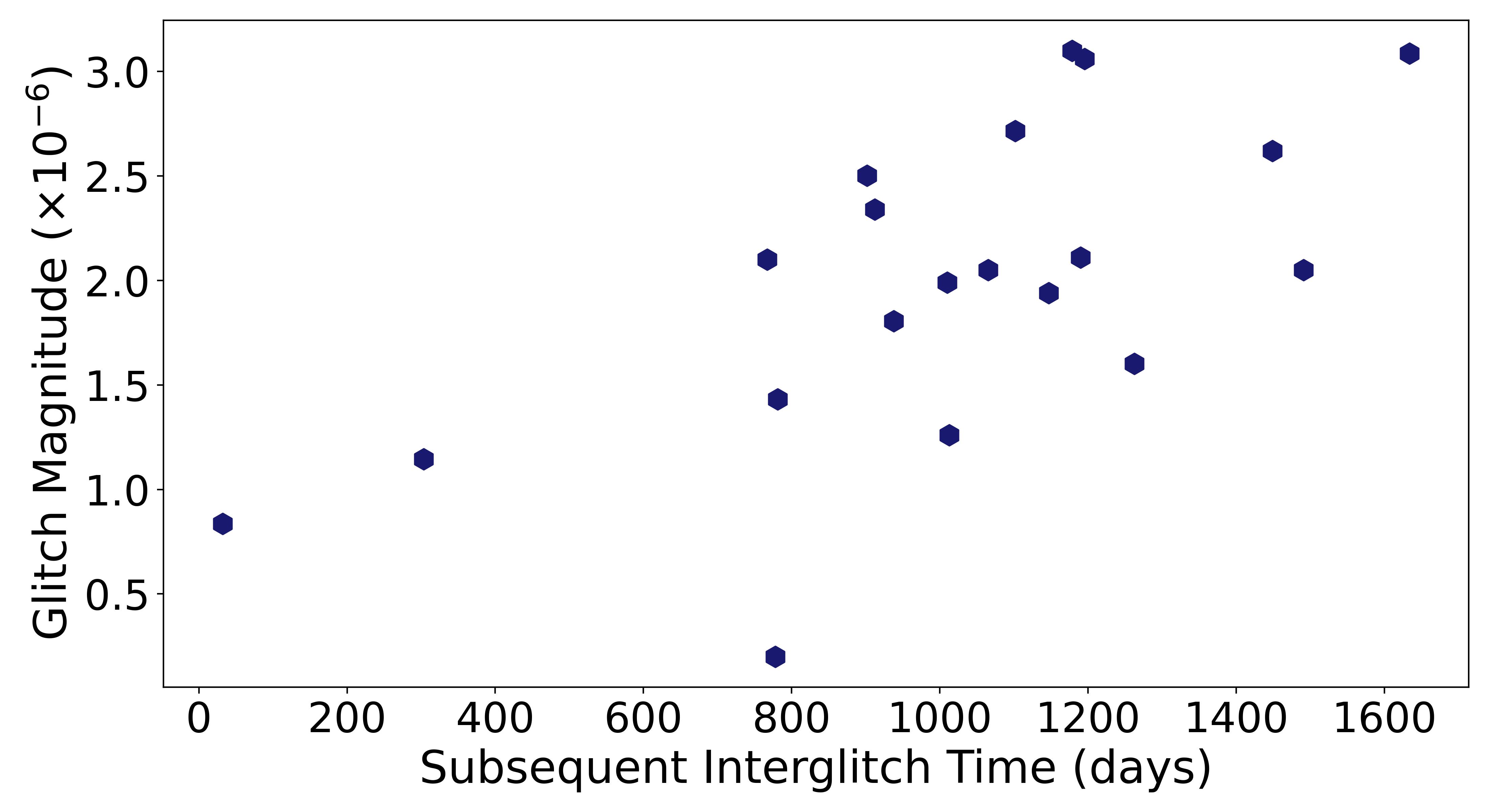}
    \caption{Glitch magnitude vs the subsequent interglitch time (in days), particularly for glitch events with large magnitudes (> 10$^{-7}$) in the Vela pulsar.}
    \label{fig: GAvsIGT}
\end{figure}

We used the reported data of 26 glitch events to investigate potential correlations between the glitch magnitude and the inter-glitch time. A significant correlation was observed between the glitch magnitude and subsequent inter-glitch time specifically for events with large magnitudes, shown in Fig.~\ref{fig: GAvsIGT}. The correlation coefficients were: Pearson's = 0.6470 (P--value = 0.0020), Spearman's = 0.5852 (P--value = 0.0067), and Kendall's = 0.4274 (P--value = 0.0086), indicating moderate to strong positive correlation. Such a correlation has only been found for PSR J0537--6910 \citep{ho20}.

\begin{figure}
    \centering
    \includegraphics[width=\linewidth]{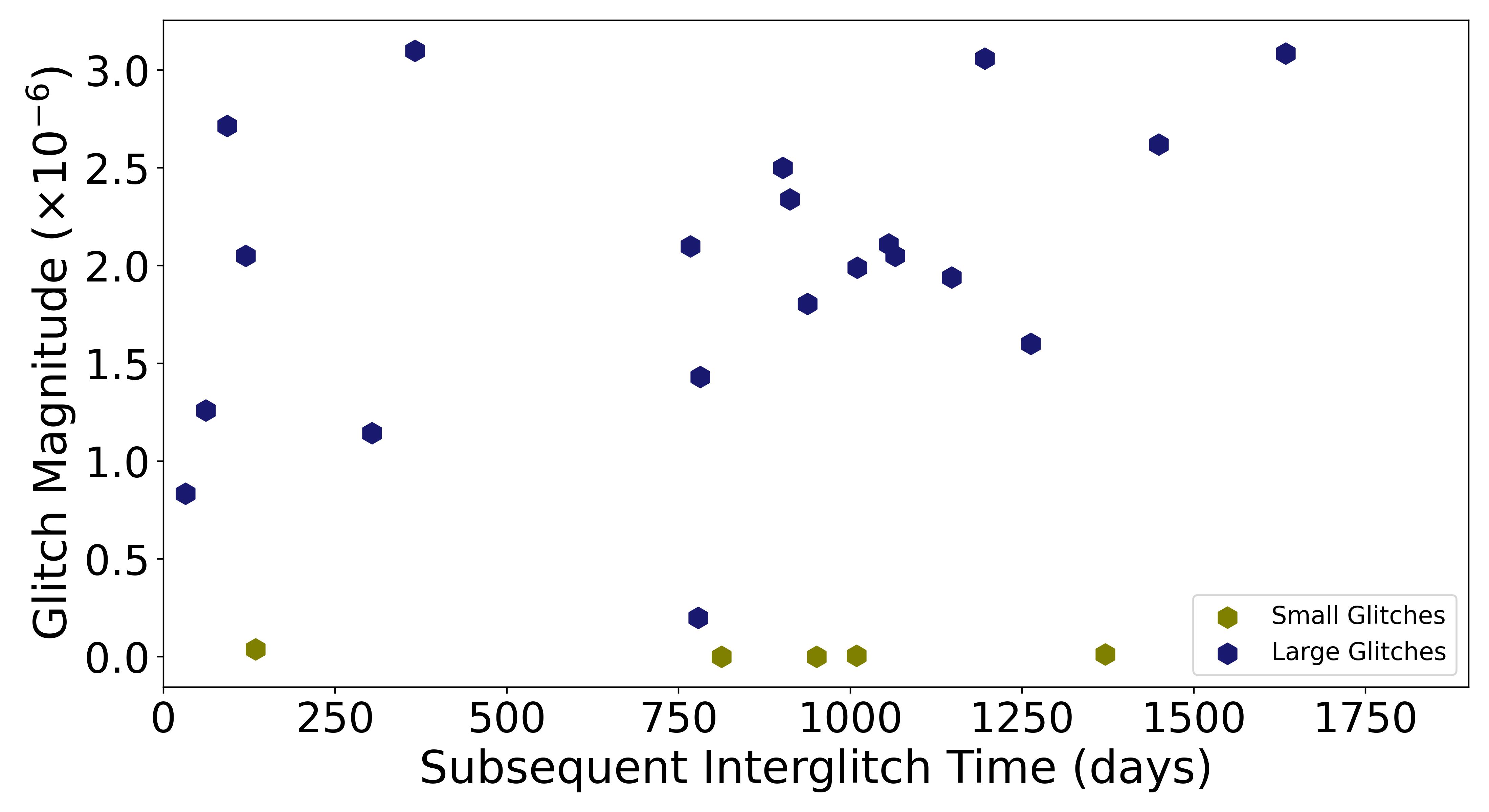}
    \caption{Glitch magnitude vs the subsequent interglitch time (in days) for all reported glitch events in the Vela pulsar.}
    \label{fig: GAvsIGT_all}
\end{figure}

This correlation between glitch magnitude and the subsequent interglitch time does not persist when small glitches are included, as illustrated in Fig.~\ref{fig: GAvsIGT_all}. The reality of reported small glitch events remains uncertain; some may be consistent with timing noise \citep{Espinoza2011, Grover2024}, although it is unlikely that all small glitches are timing noise, especially those accompanied by exponential recoveries \citep{Espinoza2021}. Therefore, the interpretation of this correlation should be approached with extreme caution. If the features and correlations shown in Figs.~\ref{fig: GAvsMJD} and~\ref{fig: GAvsIGT} persist in including all real glitches; this would have significant implications for glitch forecasting models, such as those proposed by \cite{Antonelli2022}.

We measured the Vela pulsar's braking index by estimating the frequency second derivative through linear fitting of frequency first derivative measurements just before glitch epochs, yielding a braking index value of 2.94 $\pm$ 0.55. As a caution, it may be noted that our estimate assumes almost a full recovery with stable values of the first and second spin derivatives. This value is expected to become more constrained with future observations as this would increase glitch numbers or dataset coverage. It is close to the prediction of the model that invokes the magnetic dipole radiation in a plasma-filled magnetosphere (e.g. \cite{eksi16}). Previous attempts to estimate the braking index for the Vela pulsar by different studies yielded qualitatively different results: a value of 1.4 $\pm$ 0.2 \citep{Lyne1996}, 1.7 $\pm$ 0.2~\citep{Espinoza2017} and 2.81 $\pm$ 0.12~\citep{Akbal2017}, respectively. The values in each case were different, potentially attributed to differences in techniques used to estimate spin parameters and their derivatives and assumptions.

\section{Conclusion and Future Prospects}\label{conclusion}
Focusing on the Vela pulsar's last four glitches, spanning from MJD 57651 to 60700 ($\sim$ 100 months of observations), we employed the vortex creep model to infer model parameters relevant to the post-glitch evolution. Bayesian analysis demonstrated that the recovery of the Vela glitches is well described by a hybrid model combining exponential and linear relaxations. Additionally, Bayesian inference closely predicted the observed time to the next glitch for all the glitches and suggested that the upcoming glitch is likely to occur around MJD 61377.7 (December 3, 2026), with a 68\% credible interval spanning from MJD 61249 to 61506. 

The vortex creep model provides a good description of the observed post-glitch recoveries. We further analysed the vortex residuals, defined as the difference between the model predictions and the measured spin-down rate, within the vortex bending framework. This analysis identifies damped quasi-periodic oscillations and allows estimation of the associated oscillatory parameters. We also observe an additional feature in the vortex residuals preceding the subsequent glitch, the origin of which remains unclear. These results motivate further theoretical and observational follow-up studies.

We explored several correlations and found a positive relationship between glitch magnitude and the subsequent inter-glitch interval, but only for large glitches. This suggests a possible connection between the magnitude of a glitch and the time until the next occurrence. However, this correlation does not hold when small glitches are taken into account. Consequently, a more comprehensive and robust study, with careful verification of small glitch events, is required. Furthermore, we estimated the pulsar's braking index as 2.94 $\pm$ 0.55. 

Future research includes investigating the post-glitch recovery and internal structure of other notable glitching pulsars, such as the Crab and PSR J1740--3015. To streamline the process, we are developing automated rotational parameter estimation scripts that perform the timing analysis quickly, enabling rapid model investigation and facilitating timely results. 

\section*{Acknowledgements}\label{aknowledgements}
We acknowledge the support of staff at the Radio Astronomy Centre, Ooty and the upgraded Giant Metrewave Radio Telescope during the observations. The ORT and the uGMRT are operated by the National Centre for Radio Astrophysics. This paper includes archived data obtained through the Parkes Pulsar Data archive on the CSIRO Data Access Portal (http://data.csiro.au). The Parkes radio telescope is part of the Australia Telescope, which is funded by the Commonwealth of Australia for operation as a National Facility managed by the Commonwealth Scientific and Industrial Research Organisation (CSIRO).

We acknowledge the National Supercomputing Mission (NSM) for providing computing resources of `PARAM Ganga’ at the Indian Institute of Technology Roorkee, which is implemented by C-DAC and supported by the Ministry of Electronics and Information Technology (MeitY) and Department of Science and Technology (DST), Government of India. 

\section*{Funding statement}\label{Funding statement} 
EG is supported by the Doctor Foundation of Qingdao Binhai University (No. BJZA2025025). BCJ acknowledges the support from Raja Ramanna Chair fellowship of the Department of Atomic Energy, Government of India (RRC – Track I Grant 3/3401 Atomic Energy Research 00 004 Research and Development 27 02 31 1002//2/2023/RRC/R\&D-II/13886). BCJ also acknowledges support from the Department of Atomic Energy Government of India, under project number 12-R\&D-TFR-5.02-0700. PA acknowledges the support from SERB-DST, Govt. of India, via project code CRG/2022/009359. 

\section*{Data and Codes Availability}\label{dataandcodes}

Data and codes used for this research will be provided on reasonable request.

\bibliography{references.bib}

@article{Haskell_2015,
   title={Models of pulsar glitches},
   volume={24},
   ISSN={1793-6594},
   url={http://dx.doi.org/10.1142/S0218271815300086},
   DOI={10.1142/s0218271815300086},
   number={03},
   journal={International Journal of Modern Physics D},
   publisher={World Scientific Pub Co Pte Lt},
   author={Haskell, Brynmor and Melatos, Andrew},
   year={2015},
   month={Feb},
   pages={1530008}
}

@ARTICLE{zhou_erbil_glitch_review,
       author = {{Zhou}, Shiqi and {G{\"u}gercino{\u{g}}lu}, Erbil and {Yuan}, Jianping and {Ge}, Mingyu and {Yu}, Cong},
        title = "{Pulsar Glitches: A Review}",
      journal = {Universe},
         year = 2022,
        month = dec,
       volume = {8},
       number = {12},
          eid = {641},
        pages = {641},
          doi = {10.3390/universe8120641},
archivePrefix = {arXiv},
       eprint = {2211.13885},
 primaryClass = {astro-ph.HE},
       adsurl = {https://ui.adsabs.harvard.edu/abs/2022Univ....8..641Z}}

@ARTICLE{Liu2024,
       author = {{Liu}, Y. and {Keith}, M.~J. and {Antonopoulou}, D. and {Weltevrede}, P. and {Shaw}, B. and {Stappers}, B.~W. and {Lyne}, A.~G. and {Mickaliger}, M.~B. and {Basu}, A.},
        title = "{Measuring glitch recoveries and braking indices with Bayesian model selection}",
      journal = {\mnras},
         year = 2024,
        month = jul,
       volume = {532},
       number = {1},
        pages = {859-882},
          doi = {10.1093/mnras/stae1499},
archivePrefix = {arXiv},
       eprint = {2406.09219},
 primaryClass = {astro-ph.HE},
       adsurl = {https://ui.adsabs.harvard.edu/abs/2024MNRAS.532..859L}
}

@ARTICLE{AndersonItoh1975,
   author = {{Anderson}, P.~W. and {Itoh}, N.},
    title = "{Pulsar glitches and restlessness as a hard superfluidity phenomenon}",
  journal = {Nature},
     year = 1975,
    month = jul,
   volume = 256,
    pages = {25-27},
      doi = {10.1038/256025a0},
   adsurl = {https://ui.adsabs.harvard.edu/abs/1975Natur.256...25A}
}

@ARTICLE{Alpar1984I,
       author = {{Alpar}, M.~A. and {Pines}, D. and {Anderson}, P.~W. and {Shaham}, J.},
        title = "{Vortex creep and the internal temperature of neutron stars. I - General theory}",
      journal = {\apj},
         year = 1984,
        month = jan,
       volume = {276},
        pages = {325-334},
          doi = {10.1086/161616},
       adsurl = {https://ui.adsabs.harvard.edu/abs/1984ApJ...276..325A}
}

@ARTICLE{Alpar1984II,
       author = {{Alpar}, M.~A. and {Langer}, S.~A. and {Sauls}, J.~A.},
        title = "{Rapid postglitch spin-up of the superfluid core in pulsars.}",
      journal = {\apj},
         year = 1984,
        month = jul,
       volume = {282},
        pages = {533-541},
          doi = {10.1086/162232},
       adsurl = {https://ui.adsabs.harvard.edu/abs/1984ApJ...282..533A}
}

@article{Akbal_2015,
    author = {Akbal, O. and Gügercinoğlu, E. and Şaşmaz Muş, S. and Alpar, M. A.},
    title = "{Peculiar glitch of PSR J1119−6127 and extension of the vortex creep model}",
    journal = {\mnras},
    volume = {449},
    number = {1},
    pages = {933-941},
    year = {2015},
    month = {03},
    issn = {0035-8711},
    doi = {10.1093/mnras/stv322},
    url = {https://doi.org/10.1093/mnras/stv322},
}

@ARTICLE{Erbil_2023,
       author = {{G{\"u}gercino{\u{g}}lu}, Erbil and {K{\"o}ksal}, Elif and {G{\"u}ver}, Tolga},
        title = "{On the peculiar rotational evolution of PSR B0950+08}",
      journal = {\mnras},
         year = 2023,
        month = jan,
       volume = {518},
       number = {4},
        pages = {5734-5740},
          doi = {10.1093/mnras/stac3516},
archivePrefix = {arXiv},
       eprint = {2207.04111},
 primaryClass = {astro-ph.HE},
       adsurl = {https://ui.adsabs.harvard.edu/abs/2023MNRAS.518.5734G}
}

@article{Erbil_2022,
   title={Glitches in four gamma-ray pulsars and inferences on the neutron star structure},
   volume={511},
   ISSN={1365-2966},
   url={http://dx.doi.org/10.1093/mnras/stac026},
   DOI={10.1093/mnras/stac026},
   number={1},
   journal={\mnras},
   publisher={Oxford University Press (OUP)},
   author={Gügercinoğlu, E and Ge, M Y and Yuan, J P and Zhou, S Q},
   year={2022},
   month=jan, pages={425–439} }

@ARTICLE{Radhakrishnan_Manchester_1969,
       author = {{Radhakrishnan}, V. and {Manchester}, R.~N.},
        title = "{Detection of a Change of State in the Pulsar PSR 0833-45}",
      journal = {Nature},
         year = 1969,
        month = apr,
       volume = {222},
       number = {5190},
        pages = {228-229},
          doi = {10.1038/222228a0},
       adsurl = {https://ui.adsabs.harvard.edu/abs/1969Natur.222..228R}
}

@ARTICLE{Reichley_Downs_1969,
       author = {{Reichley}, P.~E. and {Downs}, G.~S.},
        title = "{Observed Decrease in the Periods of Pulsar PSR 0833-45}",
      journal = {Nature},
         year = 1969,
        month = apr,
       volume = {222},
       number = {5190},
        pages = {229-230},
          doi = {10.1038/222229a0},
       adsurl = {https://ui.adsabs.harvard.edu/abs/1969Natur.222..229R}
}

@ARTICLE{Palfreyman2018_Vela,
       author = {{Palfreyman}, Jim and {Dickey}, John M. and {Hotan}, Aidan and {Ellingsen}, Simon and {van Straten}, Willem},
        title = "{Alteration of the magnetosphere of the Vela pulsar during a glitch}",
      journal = {Nature},
         year = 2018,
        month = apr,
       volume = {556},
       number = {7700},
        pages = {219-222},
          doi = {10.1038/s41586-018-0001-x},
       adsurl = {https://ui.adsabs.harvard.edu/abs/2018Natur.556..219P}
}

@ARTICLE{Grover2024,
       author = {{Grover}, Himanshu and {Joshi}, Bhal Chandra and {Singha}, Jaikhomba and {G{\"u}gercino{\v{g}}lu}, Erbil and {Arumugam}, Paramasivan and {Bandyopadhyay}, Debades and {Chibueze}, James O. and {Desai}, Shantanu and {Eya}, Innocent O. and {Kundu}, Anu and {Urama}, Johnson O.},
        title = "{The ORT and the uGMRT pulsar monitoring program: Pulsar timing irregularities \& the Gaussian process realisation}",
      journal = {PASA},
         year = 2024,
        month = dec,
       volume = {41},
          eid = {e102},
        pages = {e102},
          doi = {10.1017/pasa.2024.96},
archivePrefix = {arXiv},
       eprint = {2405.14351},
 primaryClass = {astro-ph.HE},
       adsurl = {https://ui.adsabs.harvard.edu/abs/2024PASA...41..102G}
}

@ARTICLE{Basu2020,
       author = {{Basu}, Avishek and {Joshi}, Bhal Chandra and {Krishnakumar}, M.~A. and {Bhattacharya}, Dipankar and {Nandi}, Rana and {Bandhopadhay}, Debades and {Char}, Prasanta and {Manoharan}, P.~K.},
        title = "{Observed glitches in eight young pulsars}",
      journal = {\mnras},
         year = 2020,
        month = jan,
       volume = {491},
       number = {3},
        pages = {3182-3191},
          doi = {10.1093/mnras/stz3230},
archivePrefix = {arXiv},
       eprint = {1911.04934},
 primaryClass = {astro-ph.HE},
       adsurl = {https://ui.adsabs.harvard.edu/abs/2020MNRAS.491.3182B}
}

@ARTICLE{ugmrt2017,
   author = { {Gupta}, Y. and {Ajithkumar}, B. and {Kale}, H.~S. and {Nayak},  S. and {Sabhapathy}, S. and {Sureshkumar}, S. and {Swami}, R.~V. and {Chengalur}, J.~N. and {Ghosh}, S.~K. and {Ishwara-Chandra}, C.~H. and {Joshi}, B.~C. and {Kanekar}, N. and {Lal}, D.~V.  and {Roy}, S. },
    title = "{The upgraded GMRT: Opening new windows on the radio Universe}",
    journal = "Current Science",
    year = 2017,
   volume = 113,
    month = jan,
    pages = {707-714},
    doi = {10.18520/cs/v113/i04/707-714}
}

@ARTICLE{Swarup1991GMRT,
       author = {{Swarup}, G. and {Ananthakrishnan}, S. and {Kapahi}, V.~K. and {Rao}, A.~P. and {Subrahmanya}, C.~R. and {Kulkarni}, V.~K.},
        title = "{The Giant Metre-Wave Radio Telescope}",
      journal = {Current Science},
         year = 1991,
        month = jan,
       volume = {60},
        pages = {95},
       adsurl = {https://ui.adsabs.harvard.edu/abs/1991CSci...60...95S}
}

@ARTICLE{Ananthakrishnan1995GMRT,
       author = {{Ananthakrishnan}, S.},
        title = "{The Giant Meterwave Radio Telescope / GMRT}",
      journal = {Journal of Astrophysics and Astronomy Supplement},
         year = 1995,
        month = jan,
       volume = {16},
        pages = {427},
       adsurl = {https://ui.adsabs.harvard.edu/abs/1995JApAS..16..427A}
}

@article{swarup1971ort,
  title={Large steerable radio telescope at Ootacamund, India},
  author={Swarup, G and Sarma, NVG and Joshi, MN and Kapahi, VK and Bagri, DS and Damle, SH and Ananthakrishnan, S and Balasubramanian, V and Bhave, SS and Sinha, RP},
  url={https://doi.org/10.1038/physci230185a0},
  DOI = {https://doi.org/10.1038/physci230185a0},
  journal={Nature Physical Science},
  volume={230},
  pages={185--188},
  year={1971}
}

@article{Naidu_2015,
   title={PONDER - A Real time software backend for pulsar and IPS observations at the Ooty Radio Telescope},
   volume={39},
   ISSN={1572-9508},
   url={http://dx.doi.org/10.1007/s10686-015-9450-5},
   DOI={10.1007/s10686-015-9450-5},
   number={2},
   journal={Experimental Astronomy},
   publisher={Springer Science and Business Media LLC},
   author={Naidu, Arun and Joshi, Bhal Chandra and Manoharan, P. K. and Krishnakumar, M. A.},
   year={2015},
   month={Mar},
   pages={319–341}
}

@ARTICLE{Hobbs2011,
       author = {{Hobbs}, G. and {Miller}, D. and {Manchester}, R.~N. and {Dempsey}, J. and {Chapman}, J.~M. and {Khoo}, J. and {Applegate}, J. and {Bailes}, M. and {Bhat}, N.~D.~R. and {Bridle}, R. and {Borg}, A. and {Brown}, A. and {Burnett}, C. and {Camilo}, F. and {Cattalini}, C. and {Chaudhary}, A. and {Chen}, R. and {D'Amico}, N. and {Kedziora-Chudczer}, L. and {Cornwell}, T. and {George}, R. and {Hampson}, G. and {Hepburn}, M. and {Jameson}, A. and {Keith}, M. and {Kelly}, T. and {Kosmynin}, A. and {Lenc}, E. and {Lorimer}, D. and {Love}, C. and {Lyne}, A. and {McIntyre}, V. and {Morrissey}, J. and {Pienaar}, M. and {Reynolds}, J. and {Ryder}, G. and {Sarkissian}, J. and {Stevenson}, A. and {Treloar}, A. and {van Straten}, W. and {Whiting}, M. and {Wilson}, G.},
        title = "{The Parkes Observatory Pulsar Data Archive}",
      journal = {\pasa},
         year = 2011,
        month = aug,
       volume = {28},
       number = {3},
        pages = {202-214},
          doi = {10.1071/AS11016},
archivePrefix = {arXiv},
       eprint = {1105.5746},
 primaryClass = {astro-ph.IM},
       adsurl = {https://ui.adsabs.harvard.edu/abs/2011PASA...28..202H}
}

@ARTICLE{DSPSR,
       author = {{van Straten}, W. and {Bailes}, M.},
        title = "{DSPSR: Digital Signal Processing Software for Pulsar Astronomy}",
      journal = {\pasa},
         year = 2011,
        month = jan,
       volume = {28},
       number = {1},
        pages = {1-14},
          doi = {10.1071/AS10021},
archivePrefix = {arXiv},
       eprint = {1008.3973},
 primaryClass = {astro-ph.IM},
       adsurl = {https://ui.adsabs.harvard.edu/abs/2011PASA...28....1V}
}

@ARTICLE{PSRCHIVE_and_PSRFITS2004,
       author = {{Hotan}, A.~W. and {van Straten}, W. and {Manchester}, R.~N.},
        title = "{PSRCHIVE and PSRFITS: An Open Approach to Radio Pulsar Data Storage and Analysis}",
      journal = {\pasa},
         year = 2004,
        month = jan,
       volume = {21},
       number = {3},
        pages = {302-309},
          doi = {10.1071/AS04022},
archivePrefix = {arXiv},
       eprint = {astro-ph/0404549},
 primaryClass = {astro-ph},
       adsurl = {https://ui.adsabs.harvard.edu/abs/2004PASA...21..302H}
}

@ARTICLE{PSRCHIVE2012,
       author = {{van Straten}, Willem and {Demorest}, Paul and {Oslowski}, Stefan},
        title = "{Pulsar Data Analysis with PSRCHIVE}",
      journal = {Astronomical Research and Technology},
         year = 2012,
        month = jul,
       volume = {9},
       number = {3},
        pages = {237-256},
archivePrefix = {arXiv},
       eprint = {1205.6276},
 primaryClass = {astro-ph.IM},
       adsurl = {https://ui.adsabs.harvard.edu/abs/2012AR&T....9..237V}
}

@article{pinta, 
title={pinta: The uGMRT data processing pipeline for the Indian Pulsar Timing Array}, 
volume={38}, 
DOI={10.1017/pasa.2021.12}, 
journal={\pasa}, 
publisher={Cambridge University Press}, author={Susobhanan, Abhimanyu and Maan, Yogesh and Joshi, Bhal Chandra and Prabu, T. and Desai, Shantanu and Nobleson, K. and Susarla, Sai Chaitanya and Girgaonkar, Raghav and Dey, Lankeswar and Batra, Neelam Dhanda and et al.}, 
year={2021}, 
pages={e017}
}

@ARTICLE{Taylor1992,
       author = {{Taylor}, J.~H.},
        title = "{Pulsar Timing and Relativistic Gravity}",
      journal = {Philosophical Transactions of the Royal Society of London Series A},
         year = 1992,
        month = oct,
       volume = {341},
       number = {1660},
        pages = {117-134},
          doi = {10.1098/rsta.1992.0088},
       adsurl = {https://ui.adsabs.harvard.edu/abs/1992RSPTA.341..117T}
}

@ARTICLE{Hobbs2006Tempo2,
       author = {{Hobbs}, G.~B. and {Edwards}, R.~T. and {Manchester}, R.~N.},
        title = "{TEMPO2, a new pulsar-timing package - I. An overview}",
      journal = {\mnras},
         year = 2006,
        month = jun,
       volume = {369},
       number = {2},
        pages = {655-672},
          doi = {10.1111/j.1365-2966.2006.10302.x},
archivePrefix = {arXiv},
       eprint = {astro-ph/0603381},
 primaryClass = {astro-ph},
       adsurl = {https://ui.adsabs.harvard.edu/abs/2006MNRAS.369..655H}
}

@ARTICLE{Edwards2006Tempo2,
       author = {{Edwards}, R.~T. and {Hobbs}, G.~B. and {Manchester}, R.~N.},
        title = "{TEMPO2, a new pulsar timing package - II. The timing model and precision estimates}",
      journal = {\mnras},
         year = 2006,
        month = nov,
       volume = {372},
       number = {4},
        pages = {1549-1574},
          doi = {10.1111/j.1365-2966.2006.10870.x},
archivePrefix = {arXiv},
       eprint = {astro-ph/0607664},
 primaryClass = {astro-ph},
       adsurl = {https://ui.adsabs.harvard.edu/abs/2006MNRAS.372.1549E}
}

@ARTICLE{Krishnakumar2021,
       author = {{Krishnakumar}, M.~A. and {Manoharan}, P.~K. and {Joshi}, B.~C. and {Girgaonkar}, R. and {Desai}, S. and {Bagchi}, M. and {Nobleson}, K. and {Dey}, L. and {Susobhanan}, A. and {Susarla}, S.~C. and {Surnis}, M.~P. and {Maan}, Y. and {Gopakumar}, A. and {Basu}, A. and {Batra}, N.~D. and {Choudhary}, A. and {De}, K. and {Gupta}, Y. and {Naidu}, A.~K. and {Pathak}, D. and {Singha}, J. and {Prabu}, T.},
        title = "{High precision measurements of interstellar dispersion measure with the upgraded GMRT}",
      journal = {\aap},
         year = 2021,
        month = jul,
       volume = {651},
          eid = {A5},
        pages = {A5},
          doi = {10.1051/0004-6361/202140340},
archivePrefix = {arXiv},
       eprint = {2101.05334},
 primaryClass = {astro-ph.HE},
       adsurl = {https://ui.adsabs.harvard.edu/abs/2021A&A...651A...5K}
}

@ARTICLE{DYNESTY2020,
       author = {{Speagle}, Joshua S.},
        title = "{DYNESTY: a dynamic nested sampling package for estimating Bayesian posteriors and evidences}",
      journal = {\mnras},
         year = 2020,
        month = apr,
       volume = {493},
       number = {3},
        pages = {3132-3158},
          doi = {10.1093/mnras/staa278},
archivePrefix = {arXiv},
       eprint = {1904.02180},
 primaryClass = {astro-ph.IM},
       adsurl = {https://ui.adsabs.harvard.edu/abs/2020MNRAS.493.3132S}
}

@BOOK{Jeffreys1961,
       author = {{Jeffreys}, Harold},
        title = "{Theory of Probability}",
        year = {reprinted 1998 (Oxford University Press, Oxford, UK, 1961)},
        edition = {3rd},
        publisher = {Oxford Classic Texts in the Physical Sciences},
}

@ARTICLE{Trotta2008,
       author = {{Trotta}, Roberto},
        title = "{Bayes in the sky: Bayesian inference and model selection in cosmology}",
      journal = {Contemporary Physics},
         year = 2008,
        month = mar,
       volume = {49},
       number = {2},
        pages = {71-104},
          doi = {10.1080/00107510802066753},
archivePrefix = {arXiv},
       eprint = {0803.4089},
 primaryClass = {astro-ph},
       adsurl = {https://ui.adsabs.harvard.edu/abs/2008ConPh..49...71T}
}

@ARTICLE{EMCEE2013,
       author = {{Foreman-Mackey}, Daniel and {Hogg}, David W. and {Lang}, Dustin and {Goodman}, Jonathan},
        title = "{emcee: The MCMC Hammer}",
      journal = {\pasp},
         year = 2013,
        month = mar,
       volume = {125},
       number = {925},
        pages = {306},
          doi = {10.1086/670067},
archivePrefix = {arXiv},
       eprint = {1202.3665},
 primaryClass = {astro-ph.IM},
       adsurl = {https://ui.adsabs.harvard.edu/abs/2013PASP..125..306F}
}

@ARTICLE{Lomb1976,
       author = {{Lomb}, N.~R.},
        title = "{Least-Squares Frequency Analysis of Unequally Spaced Data}",
      journal = {\apss},
         year = 1976,
        month = feb,
       volume = {39},
       number = {2},
        pages = {447-462},
          doi = {10.1007/BF00648343},
       adsurl = {https://ui.adsabs.harvard.edu/abs/1976Ap&SS..39..447L}
}

@ARTICLE{Scargle1982,
       author = {{Scargle}, J.~D.},
        title = "{Studies in astronomical time series analysis. II. Statistical aspects of spectral analysis of unevenly spaced data.}",
      journal = {\apj},
         year = 1982,
        month = dec,
       volume = {263},
        pages = {835-853},
          doi = {10.1086/160554},
       adsurl = {https://ui.adsabs.harvard.edu/abs/1982ApJ...263..835S}
}

@ARTICLE{VanderPlas2018,
       author = {{VanderPlas}, Jacob T.},
        title = "{Understanding the Lomb-Scargle Periodogram}",
      journal = {\apjs},
         year = 2018,
        month = may,
       volume = {236},
       number = {1},
          eid = {16},
        pages = {16},
          doi = {10.3847/1538-4365/aab766},
archivePrefix = {arXiv},
       eprint = {1703.09824},
 primaryClass = {astro-ph.IM},
       adsurl = {https://ui.adsabs.harvard.edu/abs/2018ApJS..236...16V}
}

@ARTICLE{Astropy2013,
       author = {{Astropy Collaboration} and {Robitaille}, Thomas P. and {Tollerud}, Erik J. and {Greenfield}, Perry and {Droettboom}, Michael and {Bray}, Erik and {Aldcroft}, Tom and {Davis}, Matt and {Ginsburg}, Adam and {Price-Whelan}, Adrian M. and {Kerzendorf}, Wolfgang E. and {Conley}, Alexander and {Crighton}, Neil and {Barbary}, Kyle and {Muna}, Demitri and {Ferguson}, Henry and {Grollier}, Fr{\'e}d{\'e}ric and {Parikh}, Madhura M. and {Nair}, Prasanth H. and {Unther}, Hans M. and {Deil}, Christoph and {Woillez}, Julien and {Conseil}, Simon and {Kramer}, Roban and {Turner}, James E.~H. and {Singer}, Leo and {Fox}, Ryan and {Weaver}, Benjamin A. and {Zabalza}, Victor and {Edwards}, Zachary I. and {Azalee Bostroem}, K. and {Burke}, D.~J. and {Casey}, Andrew R. and {Crawford}, Steven M. and {Dencheva}, Nadia and {Ely}, Justin and {Jenness}, Tim and {Labrie}, Kathleen and {Lim}, Pey Lian and {Pierfederici}, Francesco and {Pontzen}, Andrew and {Ptak}, Andy and {Refsdal}, Brian and {Servillat}, Mathieu and {Streicher}, Ole},
        title = "{Astropy: A community Python package for astronomy}",
      journal = {\aap},
         year = 2013,
        month = oct,
       volume = {558},
          eid = {A33},
        pages = {A33},
          doi = {10.1051/0004-6361/201322068},
archivePrefix = {arXiv},
       eprint = {1307.6212},
 primaryClass = {astro-ph.IM},
       adsurl = {https://ui.adsabs.harvard.edu/abs/2013A&A...558A..33A}
}

@ARTICLE{Baluev2008,
       author = {{Baluev}, R.~V.},
        title = "{Assessing the statistical significance of periodogram peaks}",
      journal = {\mnras},
         year = 2008,
        month = apr,
       volume = {385},
       number = {3},
        pages = {1279-1285},
          doi = {10.1111/j.1365-2966.2008.12689.x},
archivePrefix = {arXiv},
       eprint = {0711.0330},
 primaryClass = {astro-ph},
       adsurl = {https://ui.adsabs.harvard.edu/abs/2008MNRAS.385.1279B}
}

@ARTICLE{Cowan2011,
       author = {{Cowan}, Glen and {Cranmer}, Kyle and {Gross}, Eilam and {Vitells}, Ofer},
        title = "{Asymptotic formulae for likelihood-based tests of new physics}",
      journal = {European Physical Journal C},
         year = 2011,
        month = feb,
       volume = {71},
       number = {2},
          eid = {1554},
        pages = {1554},
          doi = {10.1140/epjc/s10052-011-1554-0},
archivePrefix = {arXiv},
       eprint = {1007.1727},
 primaryClass = {physics.data-an},
       adsurl = {https://ui.adsabs.harvard.edu/abs/2011EPJC...71.1554C}
}

@ARTICLE{Ganguly_Desai_2017,
       author = {{Ganguly}, Shalini and {Desai}, Shantanu},
        title = "{Statistical significance of spectral lag transition in GRB 160625B}",
      journal = {Astroparticle Physics},
         year = 2017,
        month = sep,
       volume = {94},
        pages = {17-21},
          doi = {10.1016/j.astropartphys.2017.07.003},
archivePrefix = {arXiv},
       eprint = {1706.01202},
 primaryClass = {astro-ph.IM},
       adsurl = {https://ui.adsabs.harvard.edu/abs/2017APh....94...17G}
}

@ARTICLE{Cheng1988,
       author = {{Cheng}, K.~S. and {Alpar}, M.~A. and {Pines}, D. and {Shaham}, J.},
        title = "{Spontaneous Superfluid Unpinning and the Inhomogeneous Distribution of Vortex Lines in Neutron Stars}",
      journal = {\apj},
         year = 1988,
        month = jul,
       volume = {330},
        pages = {835},
          doi = {10.1086/166517},
       adsurl = {https://ui.adsabs.harvard.edu/abs/1988ApJ...330..835C}
}

@article{Erbil_2020,
   title={The 2016 Vela glitch: a key to neutron star internal structure and dynamics},
   volume={496},
   ISSN={1365-2966},
   url={http://dx.doi.org/10.1093/mnras/staa1672},
   DOI={10.1093/mnras/staa1672},
   number={2},
   journal={\mnras},
   publisher={Oxford University Press (OUP)},
   author={Gügercinoğlu, Erbil and Alpar, M Ali},
   year={2020},
   month=jun, pages={2506–2515} }

@ARTICLE{Pizzochero2020,
       author = {{Pizzochero}, P.~M. and {Montoli}, A. and {Antonelli}, M.},
        title = "{Core and crust contributions in overshooting glitches: the Vela pulsar 2016 glitch}",
      journal = {Astronomy and Astrophysics},
         year = 2020,
        month = apr,
       volume = {636},
          eid = {A101},
        pages = {A101},
          doi = {10.1051/0004-6361/201937019},
archivePrefix = {arXiv},
       eprint = {1910.00066},
 primaryClass = {astro-ph.HE},
       adsurl = {https://ui.adsabs.harvard.edu/abs/2020A&A...636A.101P}
}

@ARTICLE{Wong2001,
       author = {{Wong}, Tony and {Backer}, D.~C. and {Lyne}, A.~G.},
        title = "{Observations of a Series of Six Recent Glitches in the Crab Pulsar}",
      journal = {\apj},
         year = 2001,
        month = feb,
       volume = {548},
       number = {1},
        pages = {447-459},
          doi = {10.1086/318657},
archivePrefix = {arXiv},
       eprint = {astro-ph/0010010},
 primaryClass = {astro-ph},
       adsurl = {https://ui.adsabs.harvard.edu/abs/2001ApJ...548..447W}
}

@ARTICLE{Shaw2018,
       author = {{Shaw}, B. and {Lyne}, A.~G. and {Stappers}, B.~W. and {Weltevrede}, P. and {Bassa}, C.~G. and {Lien}, A.~Y. and {Mickaliger}, M.~B. and {Breton}, R.~P. and {Jordan}, C.~A. and {Keith}, M.~J. and {Krimm}, H.~A.},
        title = "{The largest glitch observed in the Crab pulsar}",
      journal = {\mnras},
         year = 2018,
        month = aug,
       volume = {478},
       number = {3},
        pages = {3832-3840},
          doi = {10.1093/mnras/sty1294},
archivePrefix = {arXiv},
       eprint = {1805.05110},
 primaryClass = {astro-ph.HE},
       adsurl = {https://ui.adsabs.harvard.edu/abs/2018MNRAS.478.3832S}
}

@article{Ge_2020,
   title={Discovery of Delayed Spin-up Behavior Following Two Large Glitches in the Crab Pulsar, and the Statistics of Such Processes},
   volume={896},
   ISSN={1538-4357},
   url={http://dx.doi.org/10.3847/1538-4357/ab8db6},
   DOI={10.3847/1538-4357/ab8db6},
   number={1},
   journal={\apj},
   publisher={American Astronomical Society},
   author={Ge, M. Y. and Zhang, S. N. and Lu, F. J. and Li, T. P. and Yuan, J. P. and Zheng, X. P. and Huang, Y. and Zheng, S. J. and Chen, Y. P. and Chang, Z. and Tuo, Y. L. and Cheng, Q. and Güngör, C. and Song, L. M. and Xu, Y. P. and Cao, X. L. and Chen, Y. and Liu, C. Z. and Zhang, S. and Qu, J. L. and Bu, Q. C. and Cai, C. and Chen, G. and Chen, L. and Chen, M. Z. and Chen, T. X. and Chen, Y. B. and Cui, W. and Cui, W. W. and Deng, J. K. and Dong, Y. W. and Du, Y. Y. and Fu, M. X. and Gao, G. H. and Gao, H. and Gao, M. and Gu, Y. D. and Guan, J. and Guo, C. C. and Han, D. W. and Hao, L. F. and Huo, J. and Jia, S. M. and Jiang, L. H. and Jiang, W. C. and Jin, C. J. and Jin, J. and Jin, Y. J. and Kong, L. D. and Li, B. and Li, D. and Li, C. K. and Li, G. and Li, M. S. and Li, W. and Li, X. and Li, X. B. and Li, X. F. and Li, Y. G. and Li, Z. W. and Li, Z. X. and Liu, Z. Y. and Liang, X. H. and Liao, J. Y. and Liu, G. Q. and Liu, H. W. and Liu, X. J. and Liu, Y. N. and Lu, B. and Lu, X. F. and Luo, Q. and Luo, T. and Ma, X. and Meng, B. and Nang, Y. and Nie, J. Y. and Ou, G. and Sai, N. and Shang, R. C. and Song, X. Y. and Sun, L. and Tan, Y. and Tao, L. and Wang, C. and Wang, G. F. and Wang, J. and Wang, J. B. and Wang, M. and Wang, N. and Wang, W. S. and Wang, Y. D. and Wang, Y. S. and Wen, X. Y. and Wen, Z. G. and Wu, B. B. and Wu, B. Y. and Wu, M. and Xiao, G. C. and Xiao, S. and Xiong, S. L. and Xu, Y. H. and Yan, W. M. and Yang, J. W. and Yang, S. and Yang, Y. J. and Yang, Y. J. and Yi, Q. B. and Yin, Q. Q. and You, Y. and Yue, Y. L. and Zhang, A. M. and Zhang, C. M. and Zhang, D. P. and Zhang, F. and Zhang, H. M. and Zhang, J. and Zhang, T. and Zhang, W. C. and Zhang, W. and Zhang, W. Z. and Zhang, Y. and Zhang, Y. F. and Zhang, Y. J. and Zhang, Y. and Zhang, Z. and Zhang, Z. and Zhang, Z. L. and Zhao, H. S. and Zhao, X. F. and Zheng, W. and Zhou, D. K. and Zhou, J. F. and Zhou, X. and Zhuang, R. L. and Zhu, Y. X. and Zhu, Y.},
   year={2020},
   month=jun, pages={55} }

@ARTICLE{Erbil_2019,
    author={Gügercinoğlu, Erbil and Alpar, M Ali},
    title = "{The largest Crab glitch and the vortex creep model}",
    journal = {\mnras},
    volume = {488},
    number = {2},
    pages = {2275-2282},
    year = {2019},
    month = {07},
    issn = {0035-8711},
    doi = {10.1093/mnras/stz1831},
    url = {https://doi.org/10.1093/mnras/stz1831},
}

@ARTICLE{Alpar1989,
       author = {{Alpar}, M.~A. and {Cheng}, K.~S. and {Pines}, D.},
        title = "{Vortex Creep and the Internal Temperature of Neutron Stars: Linear and Nonlinear Response to a Glitch}",
      journal = {\apj},
         year = 1989,
        month = nov,
       volume = {346},
        pages = {823},
          doi = {10.1086/168063},
       adsurl = {https://ui.adsabs.harvard.edu/abs/1989ApJ...346..823A}
}

@ARTICLE{Flanagan1990,
       author = {{Flanagan}, Claire S.},
        title = "{Rapid recovery of the Vela pulsar from a giant glitch}",
      journal = {\nat},
         year = 1990,
        month = may,
       volume = {345},
       number = {6274},
        pages = {416-417},
          doi = {10.1038/345416a0},
       adsurl = {https://ui.adsabs.harvard.edu/abs/1990Natur.345..416F}
}

@ARTICLE{Erbil_2017aug,
       author = {{G{\"u}gercino{\u{g}}lu}, Erbil},
        title = "{Post-glitch exponential relaxation of radio pulsars and magnetars in terms of vortex creep across flux tubes}",
      journal = {\mnras},
         year = 2017,
        month = aug,
       volume = {469},
       number = {2},
        pages = {2313-2322},
          doi = {10.1093/mnras/stx985},
archivePrefix = {arXiv},
       eprint = {1701.05786},
 primaryClass = {astro-ph.HE},
       adsurl = {https://ui.adsabs.harvard.edu/abs/2017MNRAS.469.2313G}
}

@ARTICLE{Alpar1996,
       author = {{Alpar}, M.~A. and {Chau}, H.~F. and {Cheng}, K.~S. and {Pines}, D.},
        title = "{Postglitch Relaxation of the Crab Pulsar after Its First Four Major Glitches: The Combined Effects of Crust Cracking, Formation of Vortex Depletion Region and Vortex Creep}",
      journal = {\apj},
         year = 1996,
        month = mar,
       volume = {459},
        pages = {706},
          doi = {10.1086/176935},
       adsurl = {https://ui.adsabs.harvard.edu/abs/1996ApJ...459..706A}
}

@ARTICLE{zhou24,
       author = {{Zhou}, S.~Q. and {Ye}, W.~T. and {Ge}, M.~Y. and {G{\"u}gercino{\u{g}}lu}, E. and {Zheng}, S.~J. and {Yu}, C. and {Yuan}, J.~P. and {Zhang}, J.},
        title = "{A Series of (Net) Spin-down Glitches in PSR J1522{\textendash}5735: Insights from the Vortex Creep and Vortex Bending Models}",
      journal = {\apj},
         year = 2024,
        month = dec,
       volume = {977},
       number = {2},
          eid = {243},
        pages = {243},
          doi = {10.3847/1538-4357/ad938d},
archivePrefix = {arXiv},
       eprint = {2408.09204},
 primaryClass = {astro-ph.HE},
       adsurl = {https://ui.adsabs.harvard.edu/abs/2024ApJ...977..243Z}
}

@ARTICLE{zubieta25,
       author = {{Zubieta}, E. and {Garc{\'\i}a}, F. and {del Palacio}, S. and {Espinoza}, C.~M. and {Araujo Furlan}, S.~B. and {Gancio}, G. and {Lousto}, C.~O. and {Combi}, J.~A. and {G{\"u}gercino{\u{g}}lu}, E.},
        title = "{Glitch-induced pulse profile change of PSR J0742‑2822 observed from the IAR}",
      journal = {\aap},
         year = 2025,
        month = feb,
       volume = {694},
          eid = {A124},
        pages = {A124},
          doi = {10.1051/0004-6361/202452693},
archivePrefix = {arXiv},
       eprint = {2412.17766},
 primaryClass = {astro-ph.HE},
       adsurl = {https://ui.adsabs.harvard.edu/abs/2025A&A...694A.124Z}
}

@ARTICLE{liu25,
       author = {{Liu}, P. and {Yuan}, J. -P. and {Ge}, M. -Y. and {Ye}, W. -T. and {Zhou}, S. -Q. and {Dang}, S. -J. and {Zhou}, Z. -R. and {G{\"u}gercino{\u{g}}lu}, E. and {Tu}, Z. -H. and {Wang}, P. and {Li}, A. and {Li}, D. and {Wang}, N.},
        title = "{A multiband study of pulsar glitches with Fermi-LAT and Parkes}",
      journal = {\mnras},
         year = 2025,
        month = feb,
       volume = {537},
       number = {2},
        pages = {1720-1734},
          doi = {10.1093/mnras/staf101},
archivePrefix = {arXiv},
       eprint = {2408.15022},
 primaryClass = {astro-ph.HE},
       adsurl = {https://ui.adsabs.harvard.edu/abs/2025MNRAS.537.1720L}
}

@ARTICLE{Ashton2019,
       author = {{Ashton}, Gregory and {Lasky}, Paul D. and {Graber}, Vanessa and {Palfreyman}, Jim},
        title = "{Rotational evolution of the Vela pulsar during the 2016 glitch}",
      journal = {Nature Astronomy},
         year = 2019,
        month = aug,
       volume = {3},
        pages = {1143-1148},
          doi = {10.1038/s41550-019-0844-6},
archivePrefix = {arXiv},
       eprint = {1907.01124},
 primaryClass = {astro-ph.HE},
       adsurl = {https://ui.adsabs.harvard.edu/abs/2019NatAs...3.1143A}
}

@ARTICLE{Montoli2020,
       author = {{Montoli}, A. and {Antonelli}, M. and {Magistrelli}, F. and {Pizzochero}, P.~M.},
        title = "{Bayesian estimate of the superfluid moments of inertia from the 2016 glitch in the Vela pulsar}",
      journal = {\aap},
         year = 2020,
        month = oct,
       volume = {642},
          eid = {A223},
        pages = {A223},
          doi = {10.1051/0004-6361/202038340},
archivePrefix = {arXiv},
       eprint = {2005.01594},
 primaryClass = {astro-ph.HE},
       adsurl = {https://ui.adsabs.harvard.edu/abs/2020A&A...642A.223M}
}

@ARTICLE{Dunn2025,
       author = {{Dunn}, L. and {Flynn}, C. and {Bailes}, M. and {Lee}, Y.~S.~C. and {Howitt}, G. and {Melatos}, A. and {Gupta}, V. and {Mandlik}, A. and {Deller}, A.},
        title = "{First results from the UTMOST-NS pulsar timing programme}",
      journal = {\mnras},
         year = 2025,
        month = aug,
       volume = {541},
       number = {2},
        pages = {1792-1815},
          doi = {10.1093/mnras/staf1064},
archivePrefix = {arXiv},
       eprint = {2506.22697},
 primaryClass = {astro-ph.HE},
       adsurl = {https://ui.adsabs.harvard.edu/abs/2025MNRAS.541.1792D}
}

@ARTICLE{Sarkissian2017,
       author = {{Sarkissian}, John M. and {Reynolds}, John E. and {Hobbs}, George and {Harvey-Smith}, Lisa},
        title = "{One Year of Monitoring the Vela Pulsar Using a Phased Array Feed}",
      journal = {\pasa},
         year = 2017,
        month = jul,
       volume = {34},
          eid = {e027},
        pages = {e027},
          doi = {10.1017/pasa.2017.19},
archivePrefix = {arXiv},
       eprint = {1705.08355},
 primaryClass = {astro-ph.HE},
       adsurl = {https://ui.adsabs.harvard.edu/abs/2017PASA...34...27S}
}

@ARTICLE{Sarkissian2019ATel12466,
       author = {{Sarkissian}, John and {Hobbs}, George and {Reynolds}, John and {Palfreyman}, Jim and {Olney}, Steve},
        title = "{Glitch detected in the Vela Pulsar (PSR J0835-4510)}",
      journal = {The Astronomer's Telegram},
         year = 2019,
        month = feb,
       volume = {12466},
        pages = {1},
       adsurl = {https://ui.adsabs.harvard.edu/abs/2019ATel12466....1S}
}

@ARTICLE{Kerr2019ATel12481,
       author = {{Kerr}, M.},
        title = "{Fermi LAT Detection of the Recent Glitch in the Vela Pulsar (PSR J0835-4510)}",
      journal = {The Astronomer's Telegram},
         year = 2019,
        month = feb,
       volume = {12481},
        pages = {1},
       adsurl = {https://ui.adsabs.harvard.edu/abs/2019ATel12481....1K}
}

@ARTICLE{LopezArmengol2019ATel12482,
       author = {{Lopez Armengol}, F.~G. and {Lousto}, C.~O. and {del Palacio}, S. and {Garcia}, F. and {Combi}, L. and {Combi}, J.~A. and {Gancio}, G. and {Mueller}, A.~L. and {Kornecki}, P.},
        title = "{Radio observations following the recent glitch of Vela Pulsar (PSR B0833-45)}",
      journal = {The Astronomer's Telegram},
         year = 2019,
        month = feb,
       volume = {12482},
        pages = {1},
       adsurl = {https://ui.adsabs.harvard.edu/abs/2019ATel12482....1L}
}

@ARTICLE{Sosa-Fiscella2021ATel14806,
       author = {{Sosa-Fiscella}, V. and {Zubieta}, E. and {del Palacio}, S. and {Garcia}, F. and {Lopez-Armengol}, F.~A. and {Combi}, J.~A. and {Lousto}, C.~O. and {Gancio}, G. and {Combi}, L. and {Gutierrez}, E. and {Bunzel}, A. Simaz and {Hauscarriaga}, F. and {PuMA Collaboration}},
        title = "{A new Glitch in the Vela Pulsar (PSR B0833-45/PSR J0835-4510)}",
      journal = {The Astronomer's Telegram},
         year = 2021,
        month = jul,
       volume = {14806},
        pages = {1},
       adsurl = {https://ui.adsabs.harvard.edu/abs/2021ATel14806....1S}
}

@ARTICLE{Dunn2021ATel14807,
       author = {{Dunn}, L. and {Campbell-Wilson}, D. and {Flynn}, C. and {Howitt}, G. and {Lee}, Y.~S.~C. and {Melatos}, A. and {Meyers}, P. and {Bailes}, M. and {Bateman}, T. and {Day}, C. and {Deller}, A. and {Green}, A.~J. and {Gupta}, V. and {Jameson}, A. and {Lower}, M.~E. and {Mandlik}, A. and {Price}, D.~C. and {Sekhri}, R. and {Sutherland}, A. and {Torr}, G. and {Urquhart}, G.},
        title = "{Confirmation of glitch event observed in the Vela pulsar (PSR J0835-4510)}",
      journal = {The Astronomer's Telegram},
         year = 2021,
        month = jul,
       volume = {14807},
        pages = {1},
       adsurl = {https://ui.adsabs.harvard.edu/abs/2021ATel14807....1D}
}

@ARTICLE{Olney2021ATel14808,
       author = {{Olney}, Steve},
        title = "{Glitch event in the Vela pulsar (PSR J0835-4510) observed at HawkRAO}",
      journal = {The Astronomer's Telegram},
         year = 2021,
        month = jul,
       volume = {14808},
        pages = {1},
       adsurl = {https://ui.adsabs.harvard.edu/abs/2021ATel14808....1O}
}

@ARTICLE{Singha2021ATel14812,
       author = {{Singha}, Jaikhomba and {Joshi}, Bhal Chandra and {Arumugam}, Paramasivan and {Bandyopadhyay}, Debades},
        title = "{Confirmation of recent glitch in PSR J0835-4510(Vela) at 325 MHz using Ooty Radio telescope (ORT)}",
      journal = {The Astronomer's Telegram},
         year = 2021,
        month = jul,
       volume = {14812},
        pages = {1},
       adsurl = {https://ui.adsabs.harvard.edu/abs/2021ATel14812....1S}
}

@ARTICLE{Zubieta2024ATel16608_Vela,
       author = {{Zubieta}, E. and {Furlan}, S.~B. Araujo and {Palacio}, S. del and {Garcia}, F. and {Gancio}, G. and {Lousto}, C.~O. and {Combi}, J.~A. and {PuMA Collaboration}},
        title = "{Detection of a new giant glitch in the Vela Pulsar observed from the Argentine Institute of Radio astronomy}",
      journal = {The Astronomer's Telegram},
         year = 2024,
        month = may,
       volume = {16608},
        pages = {1},
       adsurl = {https://ui.adsabs.harvard.edu/abs/2024ATel16608....1Z}
}

@ARTICLE{Campbell-Wilson2024ATel16610_Vela,
       author = {{Campbell-Wilson}, D. and {Flynn}, C. and {Bateman}, T.},
        title = "{Confirmation of glitch event observed in PSR J0835-4510 (Vela pulsar)}",
      journal = {The Astronomer's Telegram},
         year = 2024,
        month = may,
       volume = {16610},
        pages = {1},
       adsurl = {https://ui.adsabs.harvard.edu/abs/2024ATel16610....1C}
}

@ARTICLE{Grover2024ATel16611_Vela,
       author = {{Grover}, Himanshu and {Krishnakumar}, M.~A. and {Joshi}, Bhal Chandra and {Arumugam}, Paramasivan},
        title = "{Vela pulsar (PSR J0835-4510) glitch observed at the Ooty Radio Telescope (ORT)}",
      journal = {The Astronomer's Telegram},
         year = 2024,
        month = may,
       volume = {16611},
        pages = {1},
       adsurl = {https://ui.adsabs.harvard.edu/abs/2024ATel16611....1G}
}

@ARTICLE{Palfreyman2024ATel16615_Vela,
       author = {{Palfreyman}, Jim},
        title = "{Epoch of the glitch of the Vela pulsar (J0835-4510) to within 4 seconds}",
      journal = {The Astronomer's Telegram},
         year = 2024,
        month = may,
       volume = {16615},
        pages = {1},
       adsurl = {https://ui.adsabs.harvard.edu/abs/2024ATel16615....1P}
}

@ARTICLE{Wang2024ATel16619_Vela,
       author = {{Wang}, Huihui and {Takata}, Jumpei and {Lin}, Lupin C.-C. and {Kong}, Albert and {Hu}, Chin-Ping and {Hui}, David and {Li}, Kwan-Lok},
        title = "{The detection of the recent glitch in the Vela pulsar by Fermi-LAT}",
      journal = {The Astronomer's Telegram},
         year = 2024,
        month = may,
       volume = {16619},
        pages = {1},
       adsurl = {https://ui.adsabs.harvard.edu/abs/2024ATel16619....1W}
}

@ARTICLE{Zubieta2025_Vela,
       author = {{Zubieta}, Ezequiel and {Missel}, Ryan and {Araujo Furlan}, Susana B. and {Lousto}, Carlos O. and {Garc{\'\i}a}, Federico and {del Palacio}, Santiago and {Gancio}, Guillermo and {Combi}, Jorge A. and {Wang}, Linwei},
        title = "{Study of the 2024 major Vela glitch at the Argentine Institute of Radioastronomy}",
      journal = {arXiv e-prints},
         year = 2025,
        month = feb,
          eid = {arXiv:2502.06704},
        pages = {arXiv:2502.06704},
          doi = {10.48550/arXiv.2502.06704},
archivePrefix = {arXiv},
       eprint = {2502.06704},
 primaryClass = {astro-ph.HE},
       adsurl = {https://ui.adsabs.harvard.edu/abs/2025arXiv250206704Z}
}

@ARTICLE{lower2020_utmost2,
       author = {{Lower}, M.~E. and {Bailes}, M. and {Shannon}, R.~M. and {Johnston}, S. and {Flynn}, C. and {Os{\l}owski}, S. and {Gupta}, V. and {Farah}, W. and {Bateman}, T. and {Green}, A.~J. and {Hunstead}, R. and {Jameson}, A. and {Jankowski}, F. and {Parthasarathy}, A. and {Price}, D.~C. and {Sutherland}, A. and {Temby}, D. and {Venkatraman Krishnan}, V.},
        title = "{The UTMOST pulsar timing programme - II. Timing noise across the pulsar population}",
      journal = {\mnras},
         year = 2020,
        month = may,
       volume = {494},
       number = {1},
        pages = {228-245},
          doi = {10.1093/mnras/staa615},
archivePrefix = {arXiv},
       eprint = {2002.12481},
 primaryClass = {astro-ph.HE},
       adsurl = {https://ui.adsabs.harvard.edu/abs/2020MNRAS.494..228L}
}

@article{Zubieta_2023,
   title={First results of the glitching pulsar monitoring programme at the Argentine Institute of Radioastronomy},
   volume={521},
   ISSN={1365-2966},
   url={http://dx.doi.org/10.1093/mnras/stad723},
   DOI={10.1093/mnras/stad723},
   number={3},
   journal={\mnras},
   publisher={Oxford University Press (OUP)},
   author={Zubieta, Ezequiel and Missel, Ryan and Sosa Fiscella, Valentina and Lousto, Carlos O and del Palacio, Santiago and López Armengol, Federico G and García, Federico and Combi, Jorge A and Wang, Linwei and Combi, Luciano and Gancio, Guillermo and Negrelli, Carolina and Gutiérrez, Eduardo M},
   year={2023},
   month=mar, pages={4504–4521} }

@ARTICLE{Espinoza2011,
       author = {{Espinoza}, C.~M. and {Lyne}, A.~G. and {Stappers}, B.~W. and {Kramer}, M.},
        title = "{A study of 315 glitches in the rotation of 102 pulsars}",
      journal = {\mnras},
         year = 2011,
        month = jun,
       volume = {414},
       number = {2},
        pages = {1679-1704},
          doi = {10.1111/j.1365-2966.2011.18503.x},
archivePrefix = {arXiv},
       eprint = {1102.1743},
 primaryClass = {astro-ph.HE},
       adsurl = {https://ui.adsabs.harvard.edu/abs/2011MNRAS.414.1679E}
}

@ARTICLE{Basu2022,
       author = {{Basu}, A. and {Shaw}, B. and {Antonopoulou}, D. and {Keith}, M.~J. and {Lyne}, A.~G. and {Mickaliger}, M.~B. and {Stappers}, B.~W. and {Weltevrede}, P. and {Jordan}, C.~A.},
        title = "{The Jodrell bank glitch catalogue: 106 new rotational glitches in 70 pulsars}",
      journal = {\mnras},
         year = 2022,
        month = mar,
       volume = {510},
       number = {3},
        pages = {4049-4062},
          doi = {10.1093/mnras/stab3336},
archivePrefix = {arXiv},
       eprint = {2111.06835},
 primaryClass = {astro-ph.HE},
       adsurl = {https://ui.adsabs.harvard.edu/abs/2022MNRAS.510.4049B}
}

@ARTICLE{ho20,
       author = {{Ho}, Wynn C.~G. and {Espinoza}, Crist{\'o}bal M. and {Arzoumanian}, Zaven and {Enoto}, Teruaki and {Tamba}, Tsubasa and {Antonopoulou}, Danai and {Bejger}, Micha{\l} and {Guillot}, Sebastien and {Haskell}, Brynmor and {Ray}, Paul S.},
        title = "{Return of the Big Glitcher: NICER timing and glitches of PSR J0537-6910}",
      journal = {\mnras},
         year = 2020,
        month = nov,
       volume = {498},
       number = {4},
        pages = {4605-4614},
          doi = {10.1093/mnras/staa2640},
archivePrefix = {arXiv},
       eprint = {2009.00030},
 primaryClass = {astro-ph.HE},
       adsurl = {https://ui.adsabs.harvard.edu/abs/2020MNRAS.498.4605H}
}

@ARTICLE{Espinoza2021,
       author = {{Espinoza}, C.~M. and {Antonopoulou}, D. and {Dodson}, R. and {Stepanova}, M. and {Scherer}, A.},
        title = "{Small glitches and other rotational irregularities of the Vela pulsar}",
      journal = {\aap},
         year = 2021,
        month = mar,
       volume = {647},
          eid = {A25},
        pages = {A25},
          doi = {10.1051/0004-6361/202039044},
archivePrefix = {arXiv},
       eprint = {2007.02921},
 primaryClass = {astro-ph.HE},
       adsurl = {https://ui.adsabs.harvard.edu/abs/2021A&A...647A..25E}
}

@INCOLLECTION{Antonelli2022,
       author = {{Antonelli}, Marco and {Montoli}, Alessandro and {Pizzochero}, Pierre M.},
        title = "{Insights Into the Physics of Neutron Star Interiors from Pulsar Glitches}",
    booktitle = {Astrophysics in the XXI Century with Compact Stars},
    editor    = {Vasconcellos, C. A. Z.},
    publisher = {World Scientific},
    address   = {Singapore},
    year      = {2022},
    pages = {219-281},
          doi = {10.1142/9789811220944_0007},
       adsurl = {https://ui.adsabs.harvard.edu/abs/2022atcc.book..219A}
}

@ARTICLE{Antonopoulou2022,
       author = {{Antonopoulou}, Danai and {Haskell}, Brynmor and {Espinoza}, Crist{\'o}bal M.},
        title = "{Pulsar glitches: observations and physical interpretation}",
      journal = {Reports on Progress in Physics},
         year = 2022,
        month = dec,
       volume = {85},
       number = {12},
          eid = {126901},
        pages = {126901},
          doi = {10.1088/1361-6633/ac9ced},
       adsurl = {https://ui.adsabs.harvard.edu/abs/2022RPPh...85l6901A}
}

@ARTICLE{eksi16,
       author = {{Ek{\c{s}}i}, K.~Y. and {Anda{\c{c}}}, I.~C. and {{\c{C}}{\i}k{\i}nto{\u{g}}lu}, S. and {G{\"u}gercino{\u{g}}lu}, E. and {Vahdat Motlagh}, A. and {K{\i}z{\i}ltan}, B.},
        title = "{The Inclination Angle and Evolution of the Braking Index of Pulsars with Plasma-filled Magnetosphere: Application to the High Braking Index of PSR J1640-4631}",
      journal = {\apj},
         year = 2016,
        month = may,
       volume = {823},
       number = {1},
          eid = {34},
        pages = {34},
          doi = {10.3847/0004-637X/823/1/34},
archivePrefix = {arXiv},
       eprint = {1603.01487},
 primaryClass = {astro-ph.HE},
       adsurl = {https://ui.adsabs.harvard.edu/abs/2016ApJ...823...34E}
}

@ARTICLE{Espinoza2017,
       author = {{Espinoza}, C.~M. and {Lyne}, A.~G. and {Stappers}, B.~W.},
        title = "{New long-term braking index measurements for glitching pulsars using a glitch-template method}",
      journal = {\mnras},
         year = 2017,
        month = apr,
       volume = {466},
       number = {1},
        pages = {147-162},
          doi = {10.1093/mnras/stw3081},
archivePrefix = {arXiv},
       eprint = {1611.08314},
 primaryClass = {astro-ph.HE},
       adsurl = {https://ui.adsabs.harvard.edu/abs/2017MNRAS.466..147E}
}

@ARTICLE{Akbal2017,
       author = {{Akbal}, O. and {Alpar}, M.~A. and {Buchner}, S. and {Pines}, D.},
        title = "{Nonlinear interglitch dynamics, the braking index of the Vela pulsar and the time to the next glitch}",
      journal = {\mnras},
         year = 2017,
        month = aug,
       volume = {469},
       number = {4},
        pages = {4183-4192},
          doi = {10.1093/mnras/stx1095},
archivePrefix = {arXiv},
       eprint = {1612.03805},
 primaryClass = {astro-ph.HE},
       adsurl = {https://ui.adsabs.harvard.edu/abs/2017MNRAS.469.4183A}
}

@ARTICLE{Lyne1996,
       author = {{Lyne}, A.~G. and {Pritchard}, R.~S. and {Graham-Smith}, F. and {Camilo}, F.},
        title = "{Very low braking index for the Vela pulsar}",
      journal = {\nat},
         year = 1996,
        month = jun,
       volume = {381},
       number = {6582},
        pages = {497-498},
          doi = {10.1038/381497a0},
       adsurl = {https://ui.adsabs.harvard.edu/abs/1996Natur.381..497L}
}

\clearpage
\onecolumn
\appendix

\section{Results of Bayesian Analysis}
\begin{figure*}
\centering
\includegraphics[width=\linewidth]{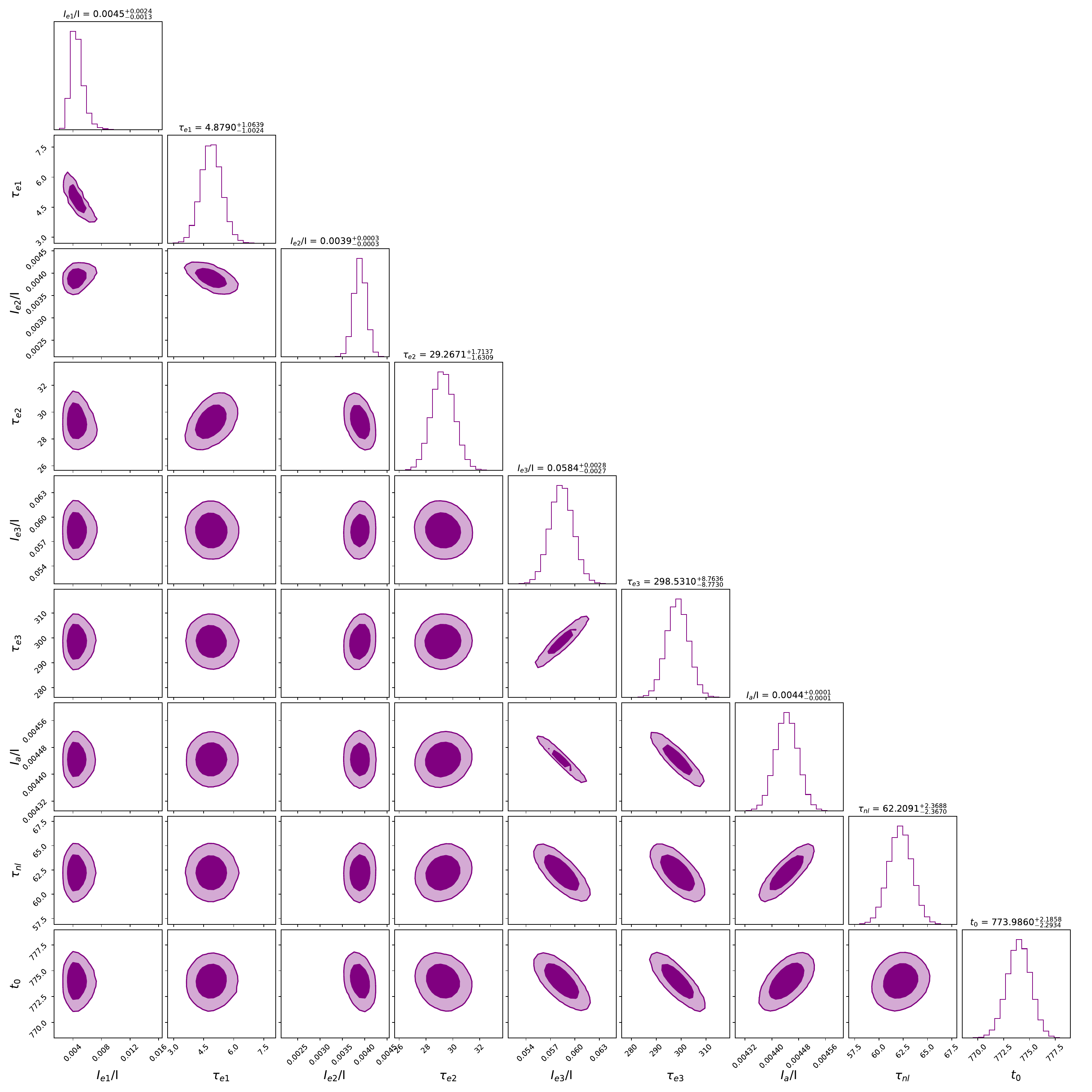}
\caption{The post-glitch recovery posteriors for PSR J0835--4510 MJD 57734 glitch with 68 and 95\% credible intervals. The symbols $I_{e1}/I$, $\tau_{e1}$,$I_{e2}/I$, $\tau_{e2}$,$I_{e3}/I$, $\tau_{e3}$ represent the fractional moment of inertia and the decay timescales for the first, second and third components of exponential recovery, respectively. And $I_{a}/I$, $\tau_{nl}$, $t_0$ denote the fractional moment of inertia, the decay timescales, and offset time associated with the linear relaxation, respectively.}
\label{posterior_Vela_57734}
\end{figure*}

\begin{figure*}
\centering
\includegraphics[width=\linewidth]{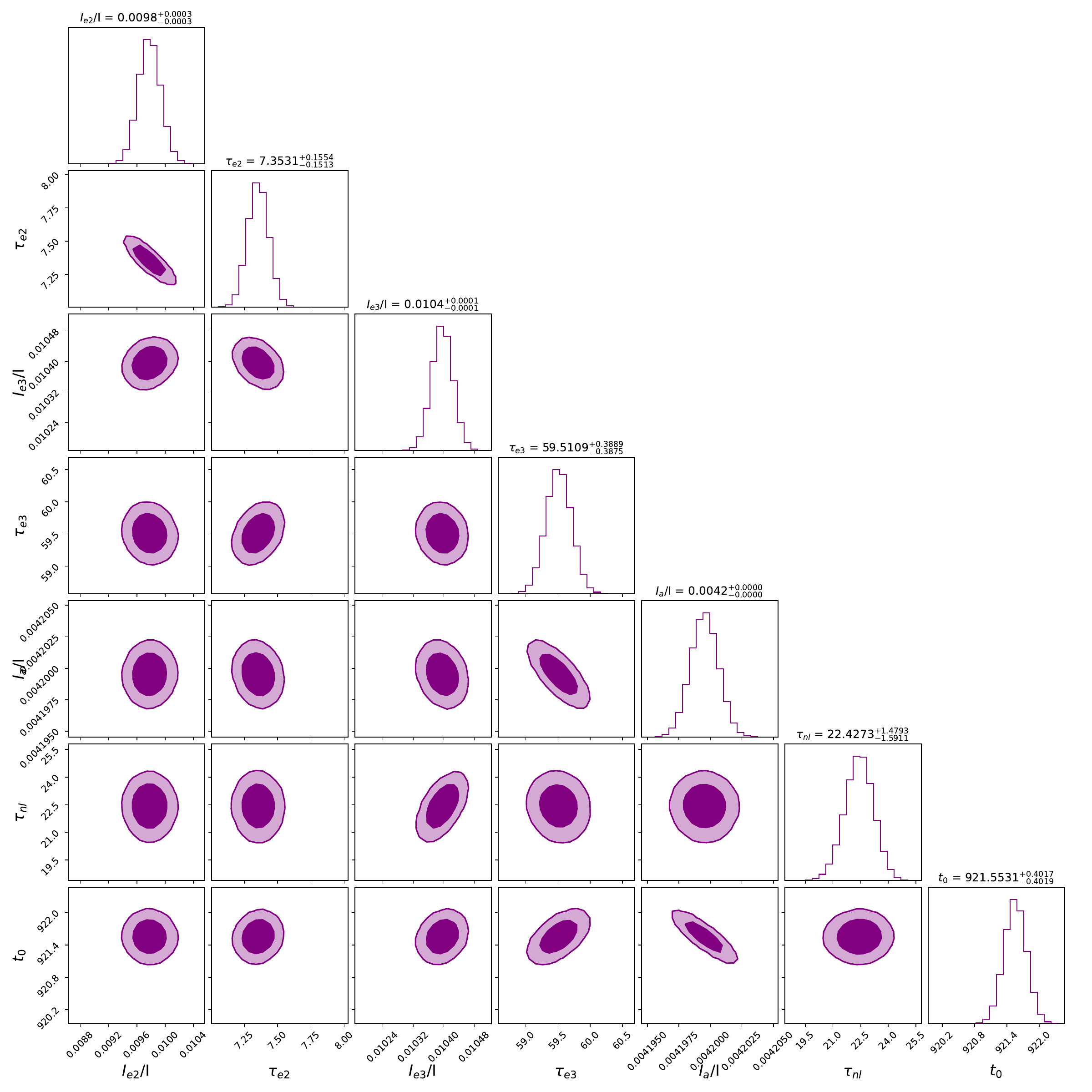}
\caption{The post-glitch recovery posteriors for PSR J0835--4510 MJD 58515 glitch with 68 and 95\% credible intervals. The symbols $I_{e2}/I$, $\tau_{e2}$,$I_{e3}/I$, $\tau_{e3}$ represent the fractional moment of inertia and the decay timescales for the second and third components of exponential recovery, respectively. And $I_{a}/I$, $\tau_{nl}$, $t_0$ denote the fractional moment of inertia, the decay timescales, and offset time associated with the linear relaxation, respectively.}
\label{posterior_Vela_58515}
\end{figure*}

\begin{figure*}
\centering
\includegraphics[width=\linewidth]{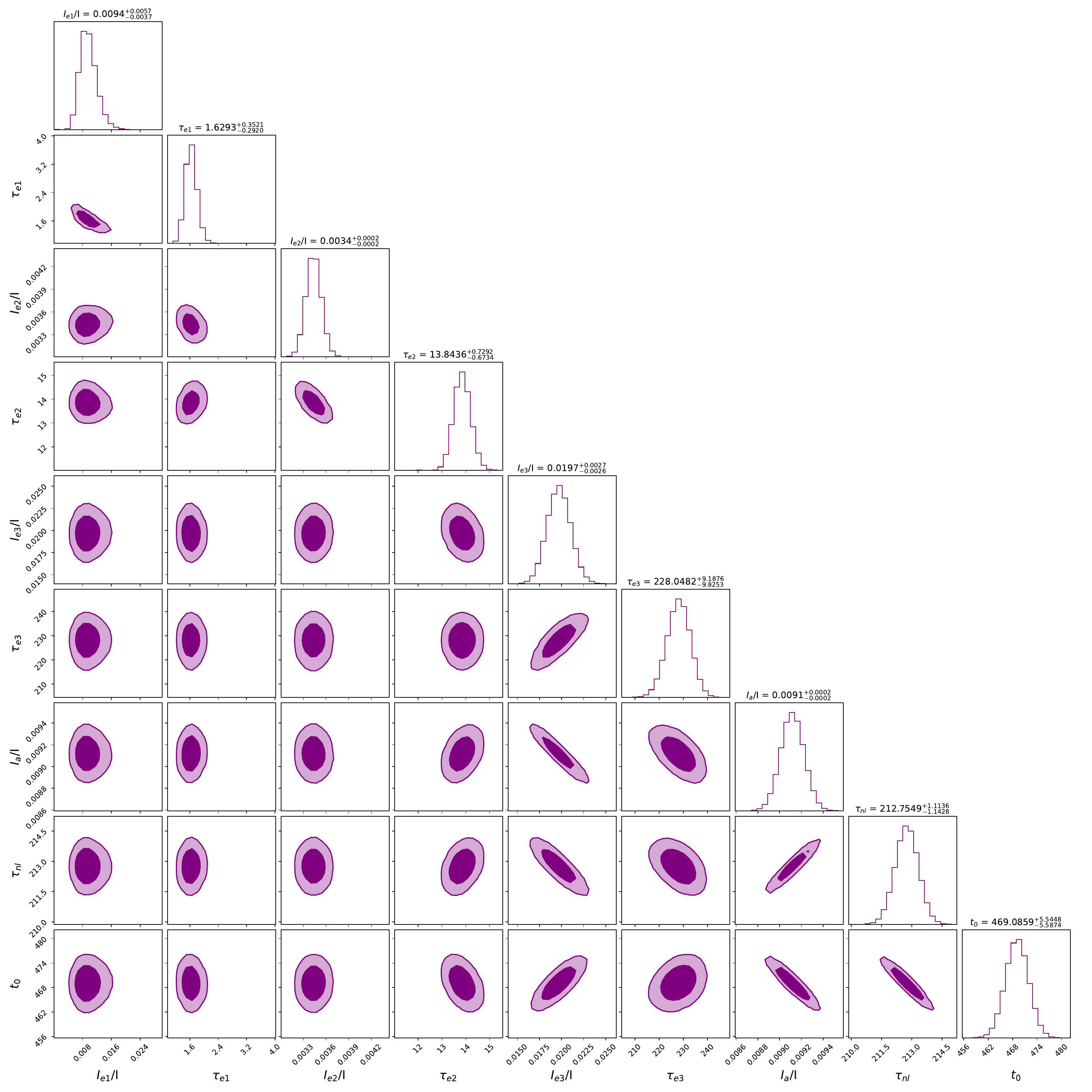}
\caption{The post-glitch recovery posteriors for PSR J0835--4510 MJD 59417 glitch with 68 and 95\% credible intervals. The symbols $I_{e1}/I$, $\tau_{e1}$,$I_{e2}/I$, $\tau_{e2}$,$I_{e3}/I$, $\tau_{e3}$ represent the fractional moment of inertia and the decay timescales for the first, second and third components of exponential recovery, respectively. And $I_{a}/I$, $\tau_{nl}$, $t_0$ denote the fractional moment of inertia, the decay timescales, and offset time associated with the linear relaxation, respectively.}
\label{posterior_Vela_59417}
\end{figure*}

\begin{figure*}
\centering
\includegraphics[width=\linewidth]{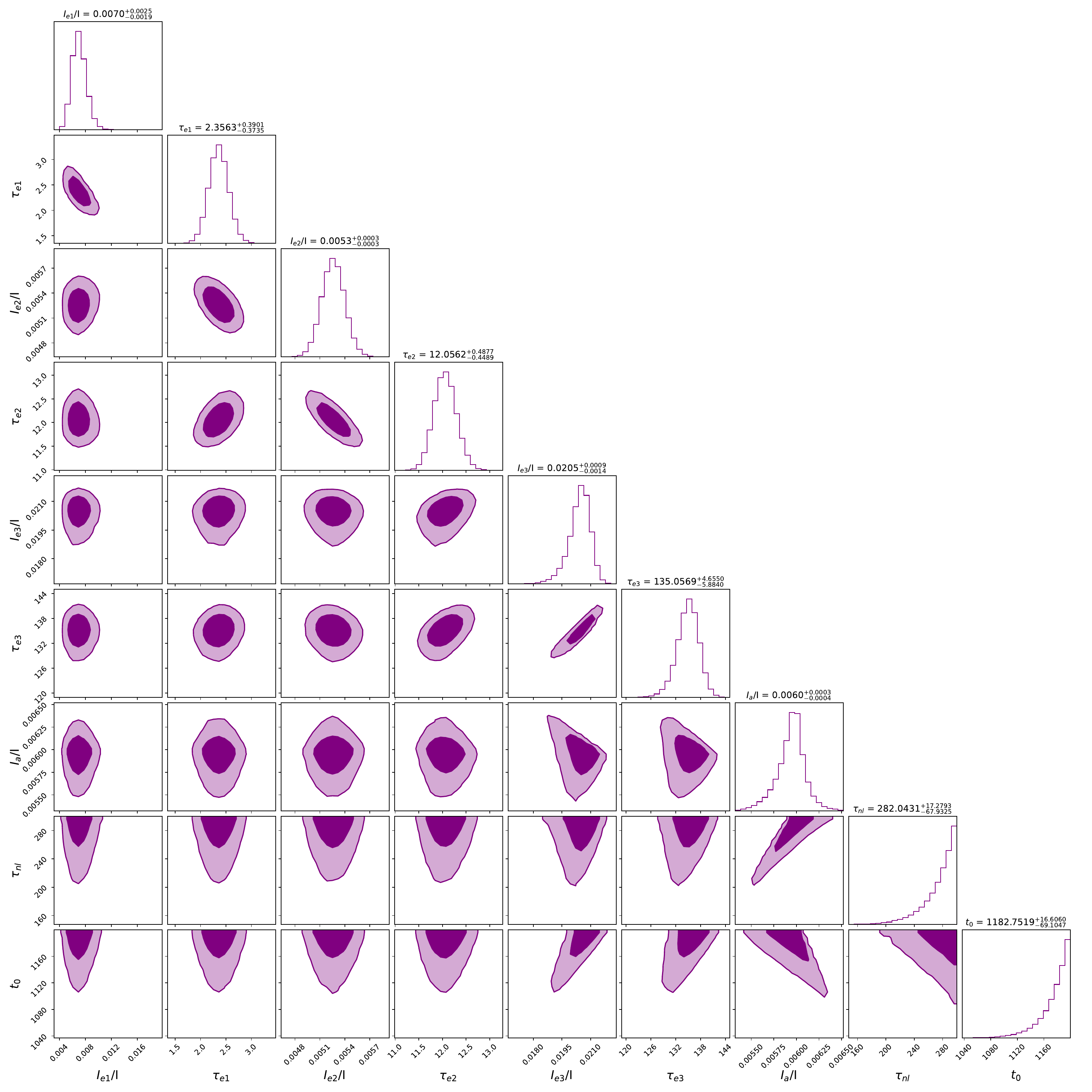}
\caption{The post-glitch recovery posteriors for PSR J0835--4510 MJD 60429 glitch with 68 and 95\% credible intervals. The symbols $I_{e1}/I$, $\tau_{e1}$,$I_{e2}/I$, $\tau_{e2}$,$I_{e3}/I$, $\tau_{e3}$ represent the fractional moment of inertia and the decay timescales for the first, second and third components of exponential recovery, respectively. And $I_{a}/I$, $\tau_{nl}$, $t_0$ denote the fractional moment of inertia, the decay timescales, and offset time associated with the linear relaxation, respectively.}
\label{posterior_Vela_60429}
\end{figure*}

\begin{figure*}
\centering
\includegraphics[width=0.45\linewidth]{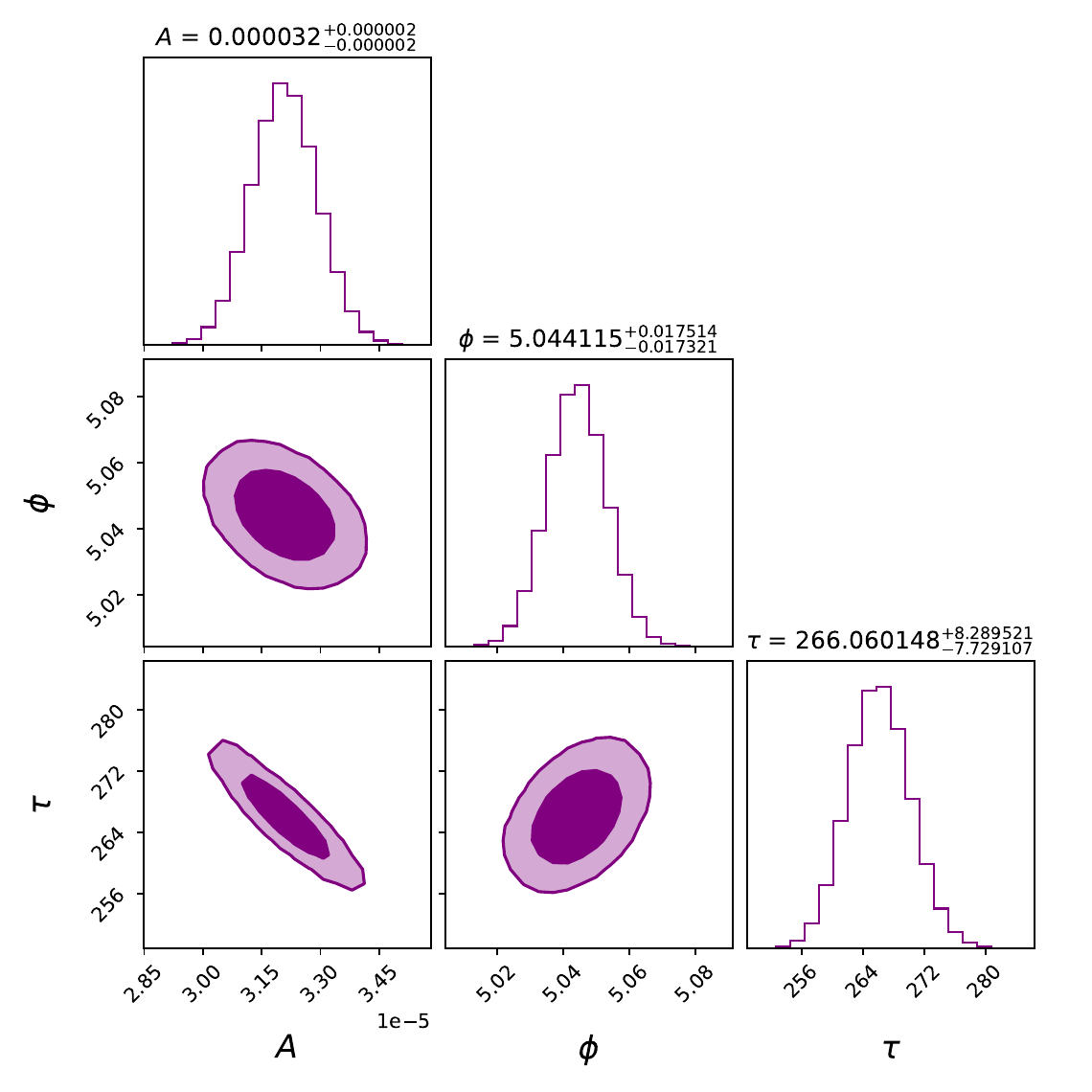}
\caption{The posteriors for the vortex residuals for PSR J0835--4510 MJD 57734 glitch with 68 and 95\% credible intervals. The symbols $A$, $\phi$, and $\tau$ represent the amplitude, phase, and decay time respectively.}
\label{posterior_residual_57734}
\end{figure*}

\begin{figure*}
\centering
\includegraphics[width=0.65\linewidth]{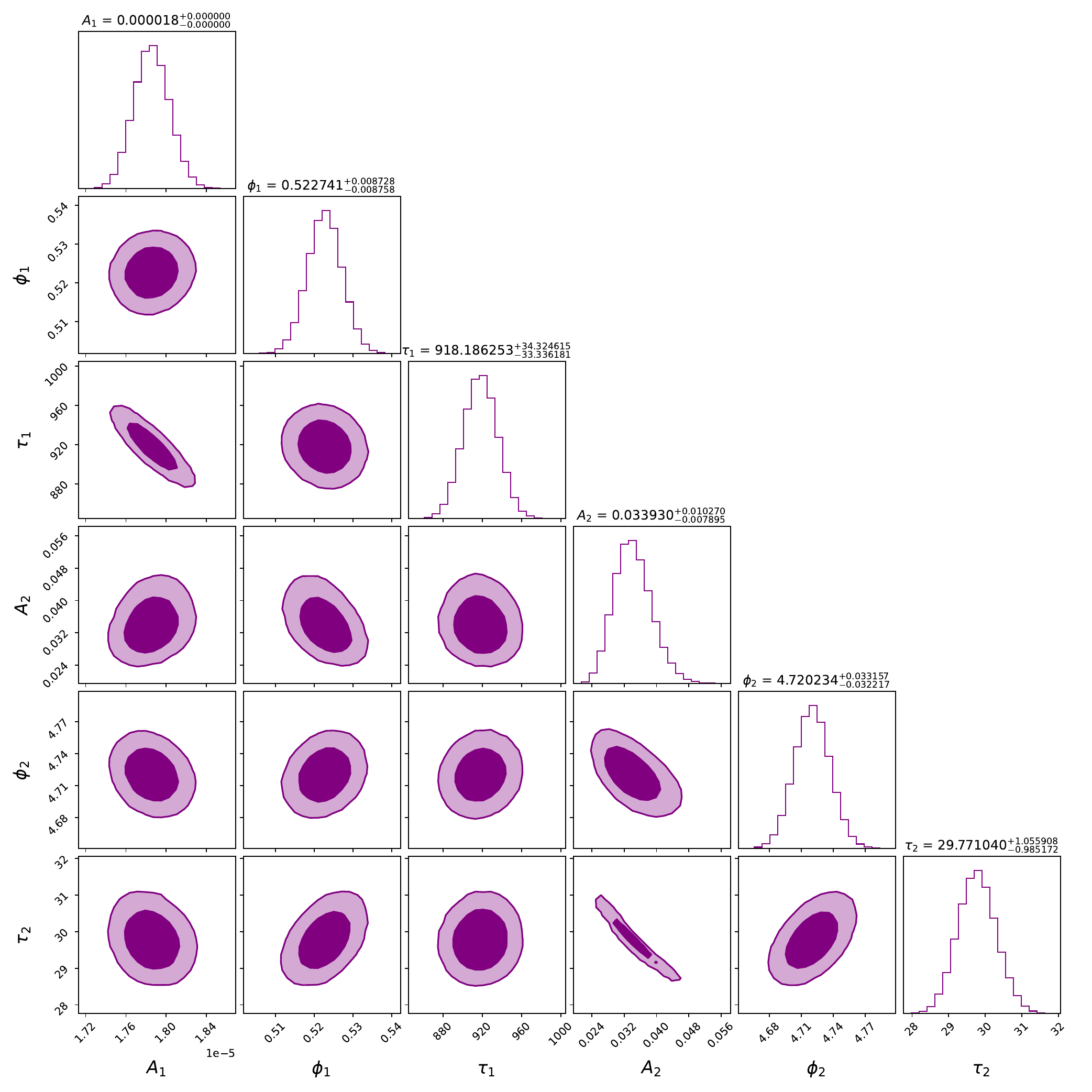}
\caption{The posteriors for the vortex residuals for PSR J0835--4510 MJD 59417 glitch with 68 and 95\% credible intervals. The symbols $A_1$, $\phi_1$, $\tau_1$ and $A_2$, $\phi_2$, $\tau_2$ denotes the amplitude, phase, decay time associated with peak1 and peak2 respectively.}
\label{posterior_residual_59417}
\end{figure*}

\noindent Here, we present the outcomes of the Bayesian model comparison and the posteriors for the parameters estimated for each post-glitch recovery using the most preferred model. The results of the Bayesian Model comparison are listed in Tables~\ref{model_select} and~\ref{model_select_vortex_residuals}. The posterior distributions are plotted with 68\% and 95\% credible intervals for all the cases given in Figs~\ref{posterior_Vela_57734}, \ref{posterior_Vela_58515}, \ref{posterior_Vela_59417}, \ref{posterior_Vela_60429}, \ref{posterior_residual_57734} and~\ref{posterior_residual_59417}.

\begin{table*}[h]
\centering
\centering
\caption{The values of $\ln(\text{BF})$ with respect to the model with the least number of free parameters for four glitches in the Vela pulsar for Models 1, 2, and 3 as described in Section~\ref{vortexCreepModel}. The bold values against a model indicate that it is the most preferred model. This preferred model has been selected based on the values of the BF.}
\label{model_select}
\renewcommand\arraystretch{2}
\setlength{\tabcolsep}{12pt}
\centering
\begin{tabular}{|c|ccc|c|ccc|}
\hline
Epoch  & Model 1a & Model 1b & Model 1c & Model 2 & Model 3a & Model 3b & Model 3c \\ \hline 
57734 & 0.0 & --14.6  & --16.6  & 126215.8 & 136246.5 & 137750.8 & \textbf{137796.8} \\
58515 & 0.0 & 28647.6 & 28644.6 & 115435.8 & 301825.3 & \textbf{309848.9} & 309846.8 \\
59417 & 0.0 & --13.9 & 71.3 & 221618.3 & 224392.2 & 224606.3 & \textbf{224627.1}  \\
60429 & 0.0 & 4451.1 & 4456.2 & 2458.6 & 14541.7 & 15608.3 & \textbf{15626.8} \\

\hline
\end{tabular}
\end{table*}

\begin{table*}[h]
\centering
\centering
\caption{The values of $\ln(\text{BF})$ with respect to the model with the least number of free parameters for the vortex residuals in the Vela pulsar corresponding to three hypotheses as described in Section~\ref{vortexCreepModel}. The bold values against a model indicate that it is the most preferred model, selected based on the values of the BF. Peak 1 or 2 `shifted' refers to measurements where the corresponding peak is assumed to originate from the preceding glitch. The BF for all the hypotheses is computed relative to the simplest hypothesis, which has the fewest free parameters and does not include any peak shift.}
\label{model_select_vortex_residuals}
\renewcommand\arraystretch{2}
\setlength{\tabcolsep}{12pt}
\centering
\begin{tabular}{|c|c|c|c|}
\hline
Epoch  & Hypothesis I & Hypothesis II & Hypothesis III \\ \hline 
57734 & 0.0 & \textbf{1283.0} & --3441.4 \\
Peak 1 Shifted & --0.1 & 886.1 & --3457.2 \\
\hline
59417 & 0.0  & \textbf{3432.4} & --14222.3 \\
Peak 1 Shifted & --4.9 & 618.1 & --23907.7 \\
Peak 2 Shifted & --37.0 & 1581.4 & --14224.4 \\
\hline
\end{tabular}
\end{table*}

\end{document}